\documentclass[authoryear]{elsarticle}
\usepackage[english]{babel}

\usepackage{natbib}

\usepackage{lipsum}
\usepackage{hyperref}
\usepackage{cleveref}
\usepackage{amsmath}
\usepackage{amsfonts}

\usepackage{amsthm}
\usepackage{framed}

\usepackage{graphicx}
\usepackage{color}
\usepackage{epsfig}
\usepackage{amssymb}
\usepackage{latexsym}
\usepackage{caption}
\usepackage{subcaption}
\allowdisplaybreaks

\newcommand{\bP}[1]{{\mathbb{P}}\left[{#1}\right]}
\newcommand{\bE}[1]{{\mathbb{E}}\left[{#1}\right]}

\newcommand{\1}[1]{{\bf 1}\left[#1\right]}       

\newcommand{\fsquare}{\vrule height6pt width7pt depth1pt}   
\newcommand{\myproof}{{\hfill \\ \bf Proof. \ }}           
\newcommand{\myendpf}{\hfill\fsquare \\[0.1in]}             
\newcommand{\myvec}[1]{{\mbox{\boldmath{$#1$}}}}

\newtheorem{theorem}{Theorem}[section]

\newtheorem{proposition}[theorem]{Proposition}
\newtheorem{lemma}[theorem]{Lemma}

\newcommand\eatpunct[1]{}

\begin{document}
\begin{frontmatter}



\title{Light traffic behavior under the power-of-two load balancing strategy: 
The case of heterogeneous servers. }


\author {  A. Izagirre$^{a}$ and A.M. Makowski$^{b}$  \\ \ \\
 $^a$Univ. of the Basque Country (UPV/EHU), Spain\\
$^b$  Department of Electrical and  Computer Engineering, and Institute for\\
 Systems Research University of Maryland, College Park, MD 20742\\
}
\address{}

\begin{abstract}
We consider a multi-server queueing system under the power-of-two policy with Poisson job arrivals, heterogeneous servers 
and a general job requirement distribution; each server operates under the first-come first-serve policy and there are no buffer constraints.
We analyze the performance of this system in light traffic by evaluating the first two light traffic derivatives of the average job response time.
These expressions point to several interesting structural features associated with server heterogeneity in light traffic:
For unequal capacities, the average job response time is seen to decrease for small values of the arrival rate,
and the more diverse  the server speeds, the greater the gain in performance.
These theoretical findings are assessed through limited simulations.

\end{abstract}


\end{frontmatter}

\section{Introduction}
\label{sec:Introduction}

Systems of parallel servers are commonly used to model resource sharing applications.
These queueing models have been adopted in classical performance studies for  supermarket cashiers, 
bank tellers and toll booths;
they have also appeared in the context of computer systems and communication networks.
A basic design issue for such systems is the {\em scheduling} of incoming jobs, usually with an eye towards 
making the average job response time as small as can be.
One possible choice is to randomly assign an incoming job to
one of the available servers, a strategy which may lead to large delays
but which has the advantage of requiring no state information.
At the other extreme,  
the join-the-shortest-queue (JSQ) policy is known to possess certain optimality properties \citep{Whitt1986},
but requires the queue length at each server to be available at the arrival epoch of every job. 

JSQ  and its variants have been extensively studied
\citep{MHB+Book} (and references therein) with
most of the work focusing on the {\em homogeneous} case 
when servers have identical service speeds and use the same service discipline.
In such cases it is known that job size variability greatly affects average
job performance  under the first-come first-serve (FCFS)
service discipline \citep[Chapter 24]{MHB+Book}.
However, the impact seems much reduced under the
processor-sharing (PS) discipline, with near-insensitivity being reported 
by  \citet{GHBSW}.

 Much work has also been done to explore the {\em trade-off} between 
the information overhead to implement job scheduling and the resulting performance.
An interesting alternative which interpolates between random assignment and JSQ
is the following policy $SQ(d)$ (for some integer $d\geq~2$):
Upon arrival,  an incoming job 
randomly selects $d$ servers from amongst the  pool of available servers. 
The JSQ policy is then applied to these $d$ servers in isolation (with a random tiebreaker) -- 
Here, shortest queue refers to the queue with the fewest jobs but other definitions (say in terms
of workload)  are possible.

This queueing system, sometimes known
as  the supermarket model, has been studied for some time now with special attention given
to the case $d=2$ (from which the terminology power-of-two derives);
see the brief historical survey in \citep[Section 1.1]{M01}.
Analysis of the supermarket model is challenging because of the coupling
between queues induced by local users of JSQ.
This is so even when jobs arrive according to a Poisson process, 
servers are identical FCFS servers,  and  job requirements are exponentially distributed.
In that setting, \citet{M01} and 
\citet{VDK96} (with $d=2$), independently,  resorted instead to studying the limiting system
obtained by letting the number of servers go to infinity.
Together their results point to
a substantial improvement in performance 
over the case $d=1$ (which corresponds to the random server assignment)
{\em without} the full overhead of {\em global} JSQ.

In view of these encouraging results it is natural to inquire whether the policy $SQ(d)$ still provides
a performance advantage when servers have {\em different} capacities.
With $d=2$,  \citet{MM} took a step in that direction: Following
the same limiting strategy as in \citep{M01,VDK96} they discuss
the average job response time for the $SQ(2)$  
model  under  {\em heterogeneous} PS servers (but with a finite number of
different server speeds), with
Poisson arrivals and a general job requirement distribution.

In this paper we consider $SQ(2)$ with {\em heterogeneous} FCFS servers, Poisson arrivals
and a general job requirement distribution.
Instead of looking at the many server asymptotics as in earlier papers,
we focus instead on the {\em light traffic} regime under a {\em fixed} number of servers; this
corresponds to the system operating with a very low traffic intensity.
Using the framework developed by  \citet{RS89}, we  compute
the first  and second light-traffic derivatives of the average job response time;
see Proposition \ref{prop:ExpectedR_k=2} in Section \ref{sec:MainResults}.
These derivatives already provide some crude {\em structural} insights into 
the impact that server heterogeneity may have on job performance; 
see Section \ref{sec:Discussion} for a short discussion.
For instance,  at least in light traffic, the more diverse  the server speeds, the greater the gain
in performance. Moreover,  job  performance in $SQ(2)$ is {\em not} monotone in the traffic intensity
(at least when this traffic intensity is small).
A quadratic polynomial \lq\lq approximation'' can be constructed on
the basis of the first two light-traffic derivatives. 
While  this {\em local} approximation cannot be accurate in moderate to heavy traffic regimes,
we nevertheless use it as a benchmark against simulations to 
illustrate the structural features revealed through  the light traffic calculations.
In Section  \ref{sec:LimitedSimulations} we further explore some of the theoretical findings with the help
of limited simulations.

The  paper is organized as follows:
The model and assumptions are introduced in Section \ref{subsec:ModelAssumptions},
and the evaluation of the first two derivatives is presented in Section \ref{subsec:FirstTwoDerivatives}.
Various comments on and implications of the results are given in Section \ref{sec:Discussion}, while
in Section  \ref{sec:LimitedSimulations} we illustrate some of the theoretical findings with the help of limited simulations.
In Section \ref{sec:ReviewLightTrafficTheory} we summarize the needed elements of the light traffic theory we use.
In  Section \ref{sec:n=0} we evaluate the light-traffic response time 
of a tagged customer, the so-called $n=0$ case in the Reiman-Simon theory.
We start the technical discussion in Section \ref{sec:AuxiliaryResult} with an auxiliary result that greatly
simplifies later computations of the first and second light-traffic derivatives.
The first derivative is computed in Section \ref{sec:n=1}.
The calculations of the second derivative start in Section \ref{sec:n=2},
and are developed through Sections \ref{sec:ProofPropExpectedR_2}--\ref{sec:ProofLemmaCase2B+2}.
Additional calculations are given in the Appendices A--E.

\section{Main results}
\label{sec:MainResults}

All random variables (rvs) under consideration in this paper are defined
on the same sufficiently large probability triple $(\Omega , {\cal A} , \mathbb{P})$; its construction
is standard and is omitted in the interest of brevity.
Probabilistic statements are made with respect to this probability measure $\mathbb{P}$,
and we denote the corresponding operator by $\mathbb{E}$.
Throughout let $\sigma$ denote an $\mathbb{R}_+$-valued rv which is 
distributed according to some probability distribution function $F: \mathbb{R}_+ \rightarrow [0,1]$,
so that $F(x) = \bP{ \sigma \leq x }$ ($x \geq 0$).
We assume at minimum that $\bE{ \sigma } < \infty $.

With any discrete set $S$  which is non-empty and finite (so $0 < |S| < \infty$), we write $U \sim \mathcal{U}(S)$
to indicate that the rv $U$ is  uniformly distributed over $S$ (under $\mathbb{P}$),
namely
\[
\bP{ U = u } = \frac{1}{|S|} ,
\quad u \in S.
\]

\subsection{Model and assumptions}
\label{subsec:ModelAssumptions}

The system comprises $K\geq 2$ parallel servers labelled $k=1, \ldots , K$.
Server $k$ has capacity $C_k $ (bytes/sec.), is attended by an infinite capacity buffer and operates
in a FCFS manner.
Jobs arrive according to a Poisson process $\{ A(t), \ t \geq 0 \}$ of rate $\lambda > 0$ with arrival epochs 
$\{ T_n, \ n=0,1, \ldots \} $ -- By convention we take $T_0=0$. For each $n=0,1, \ldots $, we refer to the job arriving at time $T_n$
as the  $n^{th}$ job; this job  brings  a random amount of work $\sigma_n$ (bytes).
Upon arrival, the  $n^{th}$ job is assigned to one of the $K$ servers
according to the  power-of-two load balancing scheme (with $d=2$):
Specifically, this incoming customer randomly selects a pair $\Sigma_n$ of distinct servers from the pool of $K$ servers. The JSQ policy is then used
in isolation with these two servers; ties are broken randomly (but other choices are possible).

As usual, the Poisson arrival process $\{ A(t), \ t \geq 0 \}$, the sequence of job requirement rvs 
$\{ \sigma_n, \ n=0,1, \ldots \}$ and the sequence
of server selection rvs  $\{ \Sigma_n, \ n=0,1, \ldots \}$ are mutually independent collections of rvs. We also assume the following:
(i) 
The rvs $\{ \sigma_n, \ n=0,1, \ldots \}$ are i.i.d. rvs distributed according to the probability distribution $F$ -- The rv $\sigma$ introduced earlier
is therefore a generic element of this sequence of i.i.d. rvs;
and 
(ii) The server selection rvs  $\{ \Sigma_n, \ n=0,1, \ldots \}$ are i.i.d. rvs, each of which is uniformly distributed
over the collection of unordered pairs drawn from $\{1, \ldots , K \}$. 
Thus, with $\mathcal{P}_2(K)$ denoting the collection of unordered pairs drawn from $\{1, \ldots , K \}$,
we have $\Sigma_n \sim \mathcal{U}(\mathcal{P}_2(K))$ with
\[
\bP{ \Sigma_n = T  } = \frac{1}{ {K \choose 2 }},
\quad 
\begin{array}{c}
T \in {\cal P}_2(K), \\
n =0,1, \ldots  \\
\end{array}
\]

Assuming the system to be initially empty (for sake of convenience),
for each $n=0,1, \ldots $,  let $R_{n,\lambda}$ denote the response time of the $n^{th}$ job
when the arrival rate is $\lambda$.
The stationary response time of a job when the arrival rate is $\lambda$ is denoted $R_\lambda$.
The existence of $R_\lambda$, possibly as an $[0,\infty]$-valued rv, can be established through 
classical semi-Markovian methods; details are omitted in the interest of brevity. 
We set
\begin{equation*}
R(\lambda ) = \bE{ R_\lambda}.
\end{equation*}
We expect $\bE{R_\lambda} < \infty$ over some non-degenerate interval $(0, \lambda^\star)$ 
for some finite $\lambda^\star > 0$, in which case $R_{n,\lambda} \Longrightarrow_n R_\lambda$
with $\Longrightarrow_n$ denoting weak convergence (also known as convergence in distribution) with $n$ going to infinity; see also
\citep[Section~2.1, Lemma~1]{M01} and \citep[Lemma~6]{MM}.
In what follows we shall not be concerned with this issue any further
since we are mainly interested in the situation where $\lambda$ is very small (vanishingly so).

\subsection{Evaluating the first two derivatives}
\label{subsec:FirstTwoDerivatives}

Light-traffic analysis considers the performance of the system for small values of the arrival rate $\lambda > 0$.
In that regime a so-called {\em light-traffic approximation} can often be constructed on the basis of 
the following Taylor series expansion argument:

Assume that for some positive integer $L$, the $L$ first derivatives of the function $\lambda \rightarrow R(\lambda)$ all exist in a neighborhood $(0,\lambda_\star)$
with $\lambda_\star > 0$.
Whenever $ 0 < x, \lambda < \lambda_\star $, Taylor's formula
\[
R(x+\lambda) 
= 
R(x) + \lambda R^{(1)} (x)
+ \frac{\lambda^2}{2!}  R^{(2)} (x) + \ldots + \frac{\lambda^L}{L!}  R^{(L)} (x) 
+ \mbox{Remainder}_L(x;\lambda)
\]
holds where we use the notation
\[
R^{(\ell)} (x)
=
\frac{d^\ell R}{d\lambda^\ell}(\lambda) \Big |_{\lambda=x} ,
\quad
 \ell =1, \ldots , L .
\]
It is not important for the discussion what is the exact form taken by the remainder term
$\mbox{Remainder}_L(x;\lambda)$.

Assume further that the limits 
\begin{equation}
R (0+)
=
\lim_{\lambda \downarrow 0} R(\lambda) 
\quad \mbox{and} \quad
R^{(\ell)} (0+)
=
\lim_{\lambda \downarrow 0} 
\frac{d^\ell R}{d\lambda^\ell}(\lambda) ,
\quad \ell =1, \ldots , L
\label{eq:LimitDerivatives}
\end{equation}
were all to exist (in $\mathbb{R}$) -- We refer to the quantities in the(\ref{eq:LimitDerivatives}) as {\em light-traffic derivatives}.
Then it is natural to use the polynomial
$R_{\rm App}(\lambda): (0,\infty) \rightarrow \mathbb{R}$ given by
\begin{equation}
R_{\rm App} (\lambda) 
= R(0+) + \lambda R^{(1)} (0+)
+ \frac{\lambda^2}{2}  R^{(2)} (0+) + \ldots + \frac{\lambda^L}{L!}  
R^{(L)} (0+) ,
\quad \lambda > 0
\label{eq:LT_Approximation}
\end{equation}
as a possible light-traffic approximation; this prompts us to write
\begin{equation}
R(\lambda) 
\simeq 
R_{\rm App} (\lambda) ,
\quad \lambda \simeq 0.
\label{eq:LT_Approximation2}
\end{equation}

The light traffic analysis presented here uses an approach proposed
by  \citet{RS89} to compute
successive light-traffic derivatives in the sense of (\ref{eq:LimitDerivatives}).
It requires that some {\em admissibility} condition be satisfied.
Following the discussion in \citep[Appendix A]{RS89} we assume 
that the generic rv $\sigma$ satisfies the condition
 \begin{equation}
 \bE{ e^{t \sigma} } < \infty
 \label{eq:ExponentialMoment}
 \end{equation}
 for some $t > 0$. This finite exponential moment condition on $F$ entails admissibility;
 it is likely stronger than needed but its purpose here is to provide a convenient framework
 where calculations can be justified. In particular it  ensures the requisite differentiability of
 $\lambda \rightarrow R(\lambda)$ where finite.
We compute the first two derivatives in light traffic; these results were announced in the conference paper
\citep{IzagirreMakowski} without proofs.


\begin{proposition}
{\sl 
Under the enforced assumptions, 
the limit $\lim_{\lambda \downarrow 0} R(\lambda ) $
exists and is given by
\begin{equation}
R(0+)
\equiv \lim_{\lambda \downarrow 0} R(\lambda )
= \frac{\Gamma}{K} \cdot \bE{ \sigma} 
\label{eq:ExpectedR_k=0}
\end{equation}
with
\begin{equation}
\Gamma = \sum_{k=1}^K \frac{1}{C_k}
\label{eq:Gamma}
\end{equation}
}
\label{prop:ExpectedR_k=0}
\end{proposition}

We now turn to the first derivative.

\begin{proposition}
{\sl  Under the enforced assumptions, the function
$\lambda \rightarrow R(\lambda)$ is differentiable in a small neighborhood
of $\lambda = 0$. Furthermore,
\begin{equation}
R^{\prime} (0+)
\equiv
\lim_{\lambda \downarrow 0} 
\frac{dR}{d\lambda} (\lambda )
=
\frac{1}{K-1}
\left( \left ( \frac{\Gamma}{K} \right )^2 
- \frac{1}{K} \sum_{k=1}^K \dfrac{1}{C_k^2} \right)
\cdot \left ( \bE{\sigma} \right )^2 .
\label{eq:ExpectedR_k=1}
\end{equation}
}
\label{prop:ExpectedR_k=1}
\end{proposition}

The third result concerns the second derivative.

\begin{proposition}
{\sl 
Under the enforced assumptions, the function
$\lambda \rightarrow R(\lambda)$ is twice differentiable 
in a small neighborhood
of $\lambda = 0$. Furthermore,
\begin{equation}
R^{\prime \prime} (0+)
\equiv
\lim_{\lambda \downarrow 0}
\frac{d^2R}{d\lambda^2} (\lambda )
=
\dfrac{2}{K^2(K-1)^2}  
\left ( 
\dfrac{\Gamma^3}{K} 
- 2\Gamma\sum_{k=1}^K \dfrac{1}{C_k^2} 
+ K\sum_{k=1}^K\dfrac{1}{C_k^3} 
\right ) 
\cdot \left ( \bE{\sigma} \right )^3.
\label{eq:ExpectedR_k=2}
\end{equation}
}
\label{prop:ExpectedR_k=2}
\end{proposition}

\section{Discussion}
\label{sec:Discussion}

\paragraph{A probabilistic interpretation}
The results of Propositions \ref{prop:ExpectedR_k=0}-\ref{prop:ExpectedR_k=2} can be expressed more compactly
with the help of the following probabilistic interpretation:
Let $X \equiv X(C_1, \ldots , C_K)$ denote a rv uniformly distributed over the set of values $\frac{1}{C_1}, \ldots , \frac{1}{C_K}$, i.e.,
$X \sim \mathcal{U}(\{ \frac{1}{C_1}, \ldots , \frac{1}{C_K} \})$ with
\[
\bP{ X = \frac{1}{C_1} }
= \ldots = \bP{ X = \frac{1}{C_K} } = \frac{1}{K}.
\]
With this notation it is easy to check that
\[
\bE{ X^p} 
= \frac{1}{K} \sum_{k=1}^K \frac{1}{C^p_k},
\quad p \geq 0.
\]
The expressions
(\ref{eq:ExpectedR_k=0}), (\ref{eq:ExpectedR_k=1}) and (\ref{eq:ExpectedR_k=2})
can now be rewritten more compactly as
\begin{equation}
R(0+) = \bE{X} \cdot \bE{\sigma} ,
\label{eq:ExpectedR_k=0Alternate} 
\end{equation}
\begin{equation}
R^{\prime} (0+)
= - \frac{1}{K-1} {\rm Var}[X] 
\cdot \left (  \bE{\sigma } \right )^2
\label{eq:ExpectedR_k=1Alternate} 
\end{equation}
and
\begin{eqnarray}
R^{\prime \prime} (0+)
&=&
\dfrac{2}{(K-1)^2}  
\left (
\left ( \bE{X} \right )^3 - 2 \bE{X} \cdot \bE{X^2}  + \bE{X^3}
\right )   \left ( \bE{\sigma} \right )^3
\nonumber \\
&=&
\dfrac{2}{(K-1)^2}  
\left (
 \bE{X^3} - \left ( \bE{X} \right )^3  - 2 \bE{X} \cdot \rm{Var}[ X ]
\right )   \left ( \bE{\sigma} \right )^3 ,
\label{eq:ExpectedR_k=2Alternate}  
\end{eqnarray}
respectively.

\paragraph{Equal capacities}
From (\ref{eq:ExpectedR_k=1Alternate}) it follows that $R^{\prime} (0+) \leq 0$, with $R^{\prime} (0+) = 0$ if and only if
${\rm Var}[X]  =~0$, or equivalently, $C_1 = \ldots = C_K $.
In that case all $K$ servers have the same capacity, and we also have
$R^{\prime \prime} (0+)  = 0 $, whence
\[
R(\lambda) = \frac{\bE{\sigma} }{C}+ o(\lambda^2)
\]
assuming the existence of a third derivative (via either the Lagrange or Cauchy form of the remainder).

\paragraph{Unequal capacities}
When the capacities are different, then $R^{\prime} (0+) < 0$ and $R(\lambda) $ is
{\em decreasing} for small values of $\lambda $.
This  is a somewhat unexpected finding because most queueing systems are \lq\lq monotone''
in the sense  that  increasing the traffic intensity $\lambda$
results in an increase in a performance metric such as the average job response time.

This  fact can be explained as follows:
On the average, a job entering an empty system experiences a response time given by $R(0+)$
since the scheduling policy $SQ(2)$ assigns it to {\em any} of the $K$ servers with probability $\frac{1}{K}$.
However, when the servers have different capacities, the assigned server may not have been the fastest,
therefore making it possible for subsequent jobs to be served by faster servers by the luck of the draw. 
This will result in a decrease in the average job response time if the traffic intensity increases slightly but still allows 
for some faster  server to be available with some non-negligible probability.

\paragraph[How much of a decrease?]{How much of a decrease?\eatpunct} 
We see from (\ref{eq:ExpectedR_k=1Alternate}) that the decrease in the average job response time
will be more pronounced the larger the variance $\rm{Var}[X]$ of the rv $X$. 
It is therefore natural to wonder which set of capacity values $C_1, \ldots , C_K$
yield the largest value for this variance $\rm{Var}[X]$ under a given value for $\bE{X}$,
say $\bE{X} = \frac{\Gamma}{K}$ for some $\Gamma > 0$.
As we assess the range of $\rm{Var}[X]$ under this constraint on $\bE{X}$
in Appendix A,  we conclude  that
\begin{equation}
0 \leq \rm{Var}[X] < \frac{\Gamma^2}{K} - \left ( \frac{\Gamma}{K} \right )^2 
= (K-1) \left ( \frac{\Gamma}{K} \right )^2
\quad \mbox{with $\bE{X} = \frac{\Gamma}{K}$}.
\label{eq:VAR}
\end{equation}

As mentioned earlier, the lower bound is achievable by the vector of capacities given by
\begin{equation}
\myvec{C}_\star = \left ( \frac{K}{\Gamma} , \ldots , \frac{K}{\Gamma} \right ).
\label{eq:BalancedCapacities}
\end{equation}
While the upper bound is {\em not} achievable by any vector of capacities satisfying the constraint, it is however
tight in the following sense: 
For each $k=1, \ldots , K$, let the vector $\myvec{e}_k$ 
denote the $K$-dimensional vector $(\delta_{k\ell})$ with all zero entries except in the $k^{th}$ position where it is one. 
The vectors of capacities given by
\begin{equation}
\myvec{C}_{k,a} 
= \frac{1}{a\Gamma} \myvec{e}_k + \frac{K-1}{(1-a) \Gamma} \sum_{\ell=1, \ \ell \neq k}^K \myvec{e}_\ell,
\quad
\begin{array}{c}
k=1, \ldots , K \\
0 < a < 1 \\
\end{array}
\label{eq:C_ak}
\end{equation}
can approach the upper bound value arbitrarily close by letting $a$ go to $1$; this is shown in 
Appendix A.

The lower bound is implemented by the most balanced capacity assignment (\ref{eq:BalancedCapacities})
under the constraint $\bE{X} = \frac{\Gamma}{K}$, whereas the upper bound is achieved, 
albeit  asymptotically, by capacity assignments 
(\ref{eq:C_ak}) that are as imbalanced as they can be under the constraint. In the limit these assignments
correspond to $K-1$ servers that are infinitely fast with the remaining \lq\lq slow" one with finite capacity.

\paragraph{Only $\bE{ \sigma }$ matters}
The two first derivatives at $\lambda~=~0+$
depend only on the first moment of $\sigma$, and could be read as a form
of insensitivity in light traffic. This is in sharp contrast with other systems where the first light-traffic derivative
depends on $\bE{\sigma^2}$, e.g.,  $M|G|1$-like queues \citep{RS88} 
and the discriminatory processor sharing model  \citep{IAM}.
This is rather unexpected because the variance of 
 $\sigma$ is known to be a key factor in shaping JSQ performance 
 with homogeneous servers under FCFS scheduling
 \citep[Chapter 24]{MHB+Book}. See next item for a possible explanation.

\paragraph{FCFS vs. PS}
Proposition \ref{prop:ExpectedR_k=2}
was established under the assumption that the servers operate
under the FCFS discipline.
It is easy to see that both 
(\ref{eq:ExpectedR_k=0}) and (\ref{eq:ExpectedR_k=1}) (but not
(\ref{eq:ExpectedR_k=2})) are still valid if the servers all use the PS discipline: This is because in the cases
$n=0$ and $n=1$ the tagged job will not share a server with another job under either discipline;
see Section \ref{sec:n=0} and Section \ref{sec:n=1}. However, this changes for the case $n=2$ that involves three customers.
That the variability of $\sigma$ seems to play little role in light traffic is therefore
consistent with the  aforementioned fact that performance under the PS discipline
is nearly insensitive to service variability \citep{GHBSW}.

\section{Limited simulations}
\label{sec:LimitedSimulations}


\begin{figure*}
\centering
\begin{subfigure}[b]{.24\textwidth}
  \includegraphics[width=0.99\textwidth]{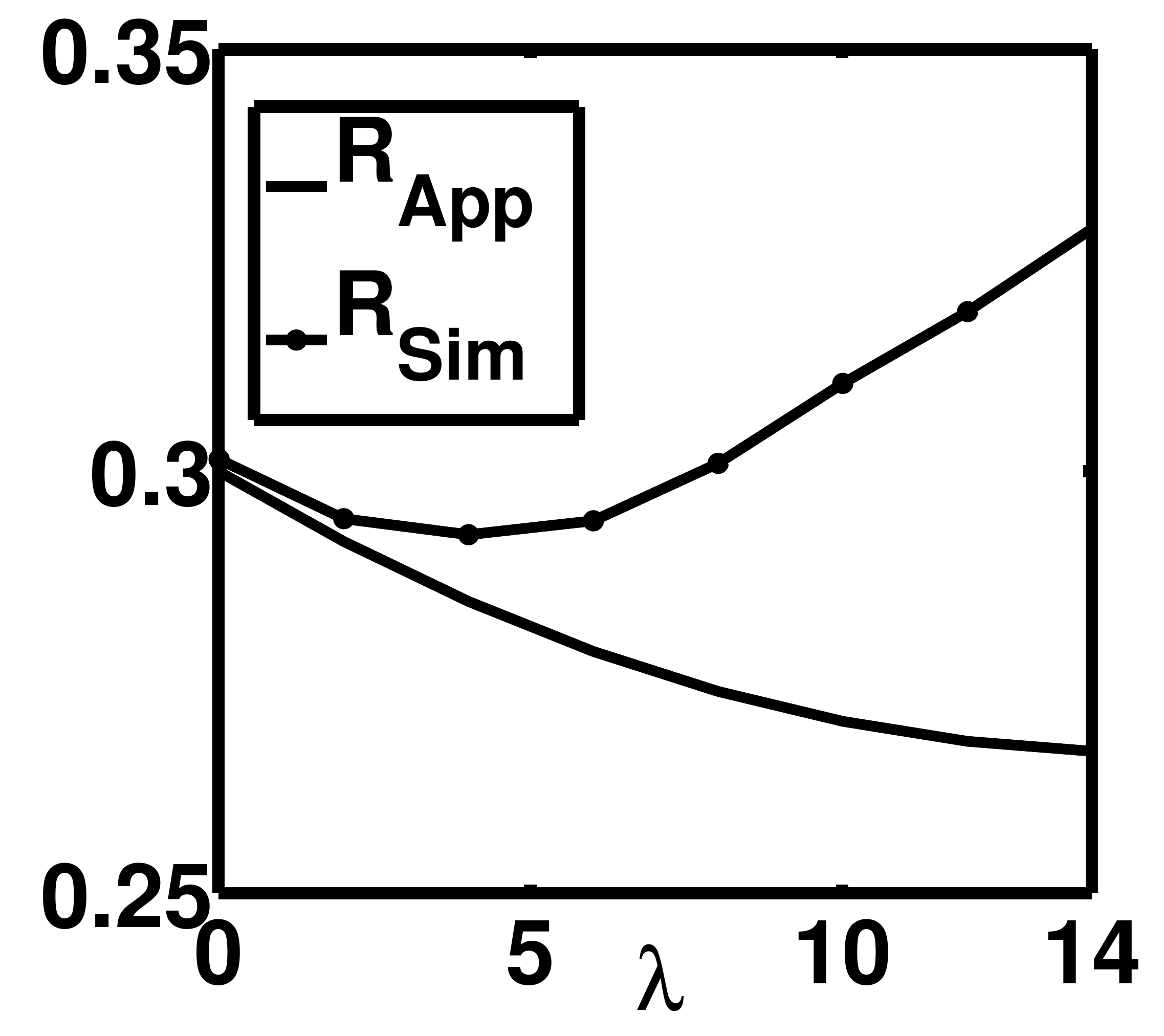}
  \caption{Hyper-exponential}
      \label{SQ2 fig:Sce1_hyper}
  	         \end{subfigure}    
		\begin{subfigure}[b]{.24\textwidth}
  \includegraphics[width=0.92\textwidth]{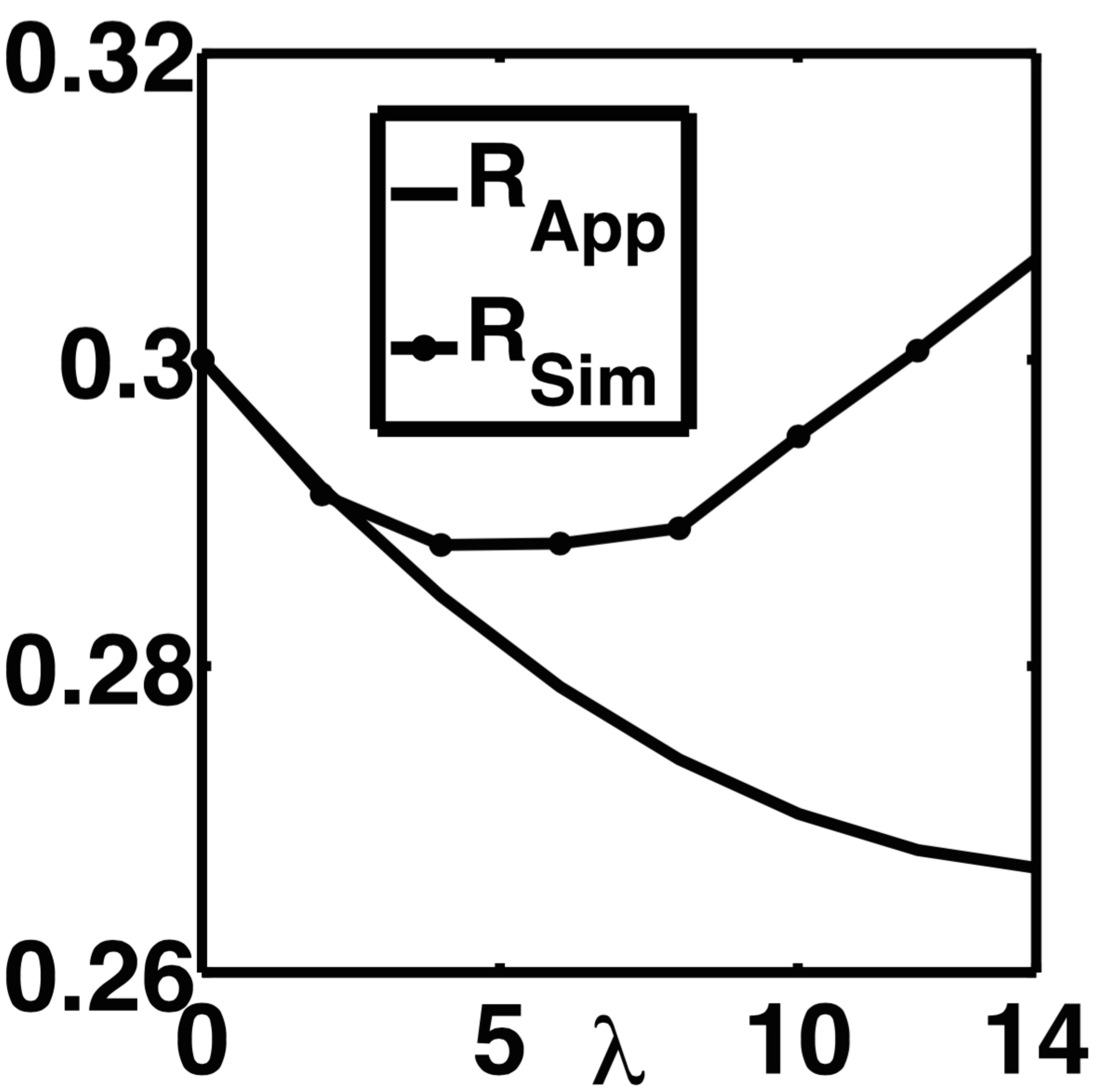}
  \caption{Exponential}
      \label{SQ2 fig:Sce1_exp}
  	         \end{subfigure}       
	           		\begin{subfigure}[b]{.24\textwidth}
  \includegraphics[width=0.99\textwidth]{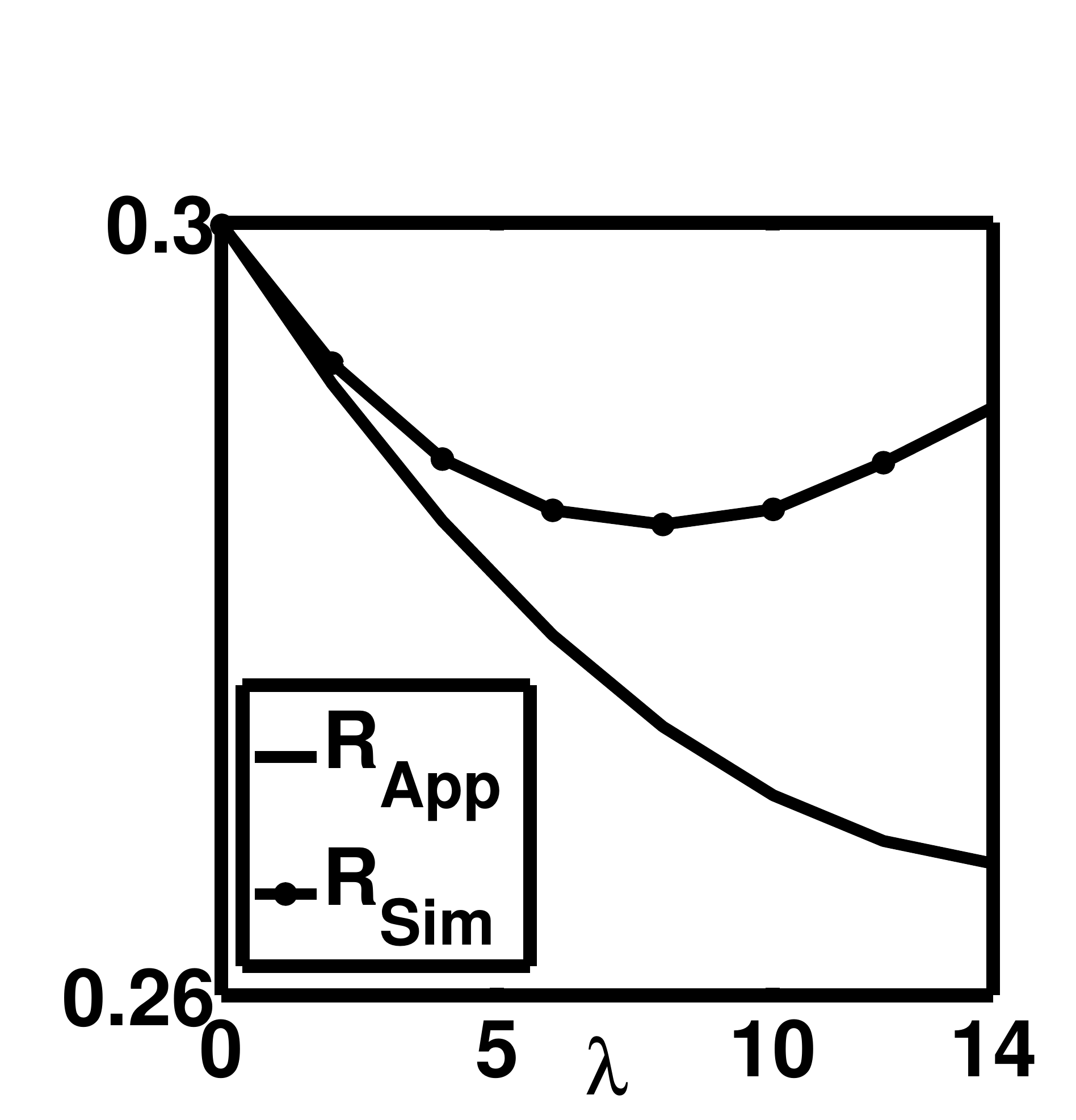}
  \caption{Weibull}
        \label{SQ2 fig:Sce1_weibull}
  	         \end{subfigure}	       
		\begin{subfigure}[b]{.24\textwidth}
  \includegraphics[width=0.99\textwidth]{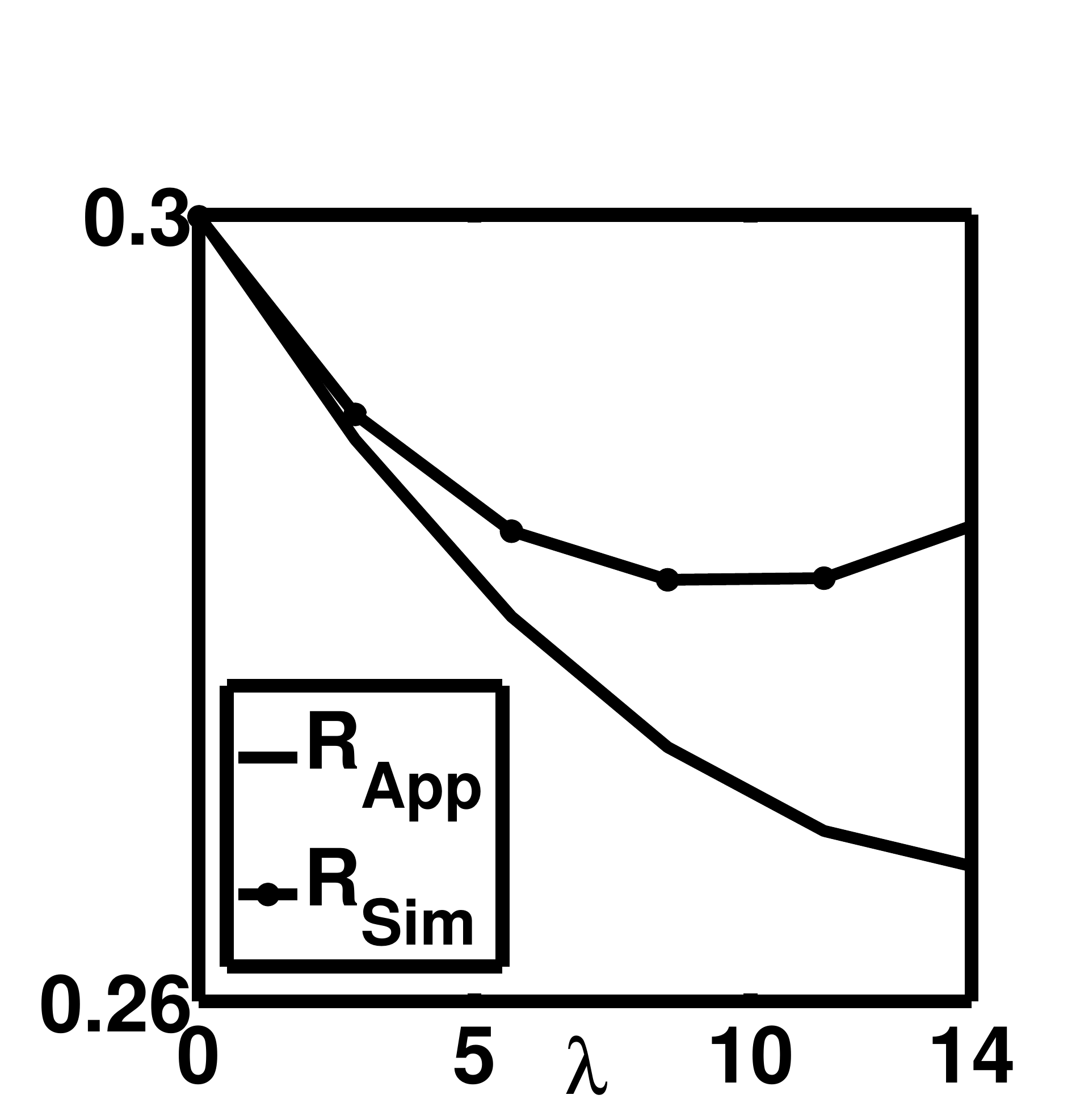}
  \caption{Deterministic}
        \label{SQ2 fig:Sce1_deter}
  	         \end{subfigure}
	         \caption{Scenario 1}
	         \label{SQ2 fig:Sce1}
\end{figure*}

As explained in Section \ref{subsec:FirstTwoDerivatives} the second order polynomial
\begin{equation}
R_{\rm App} (\lambda) =  R(0+) + \lambda R^{\prime } (0+)  + \frac{\lambda^2}{2} R^{\prime \prime} (0+),
\quad \lambda  \geq 0
\label{SQ2 eq:LT_App_degree2}
\end{equation}
can be used as a local approximation to $R(\lambda)$ for small $\lambda$.
As already pointed out by  \citet{RS88,RS89}, without additional information (e.g., heavy traffic information),
we should not expect $R_{\rm App}(\lambda)$ to act as an accurate proxy for $R(\lambda)$ in medium to heavy traffic.
This lack  of accuracy is certainly apparent in the simulation results reported  below.

We have carried out 
simulations for different distributions of $\sigma$,
all with unit mean, namely hyperexponential (obtained by mixing  the exponential rvs
${\rm Exp}(1/2)$ and ${\rm Exp}(2)$ with probability $1/3$ and $2/3$, respectively), exponential ${\rm Exp(1)}$
(of parameter $1$),
Weibull (with shape parameter $2$ and scale parameter $\Gamma(3/2)^{-1}$) and deterministic.
The simulation results are based on averaging $10$ runs
with each run comprising  $10^5$ busy periods. 
A busy period is defined as the interval of time between two consecutive time epochs when the system becomes empty, 
such points being regenerative points for the  stochastic process of interest.
We have verified that the simulation results obtained for a system with $K=100$ homogeneous servers 
and exponential service requirements agree with those given by \citet[Table~1]{M01}.
\begin{figure*}
\centering
\begin{subfigure}[b]{.24\textwidth}
  \includegraphics[width=0.99\textwidth]{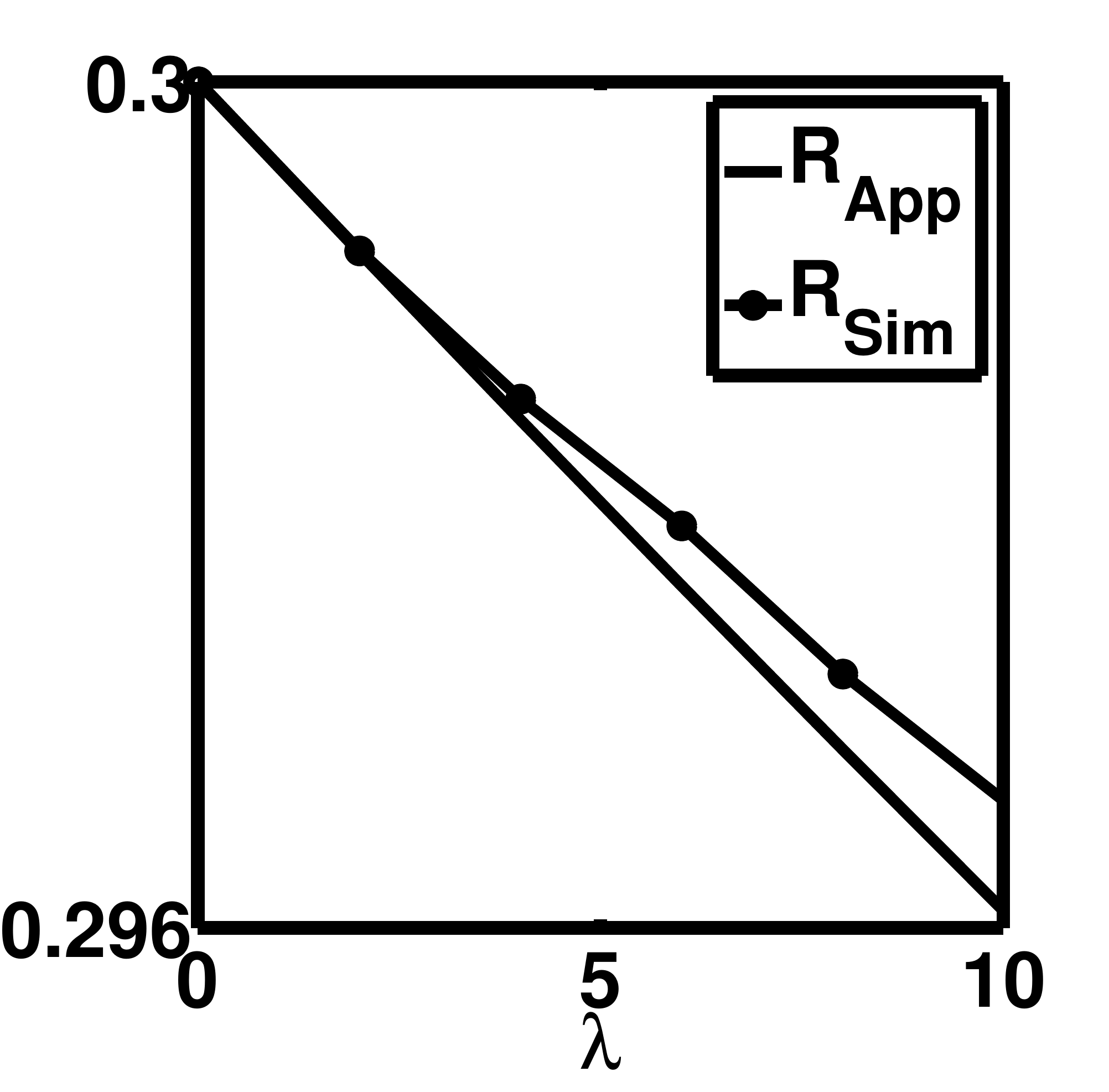}
  \caption{Hyper-exponential}
      \label{SQ2 fig:Sce2_hyper}
  	         \end{subfigure}    
		\begin{subfigure}[b]{.24\textwidth}
  \includegraphics[width=0.99\textwidth]{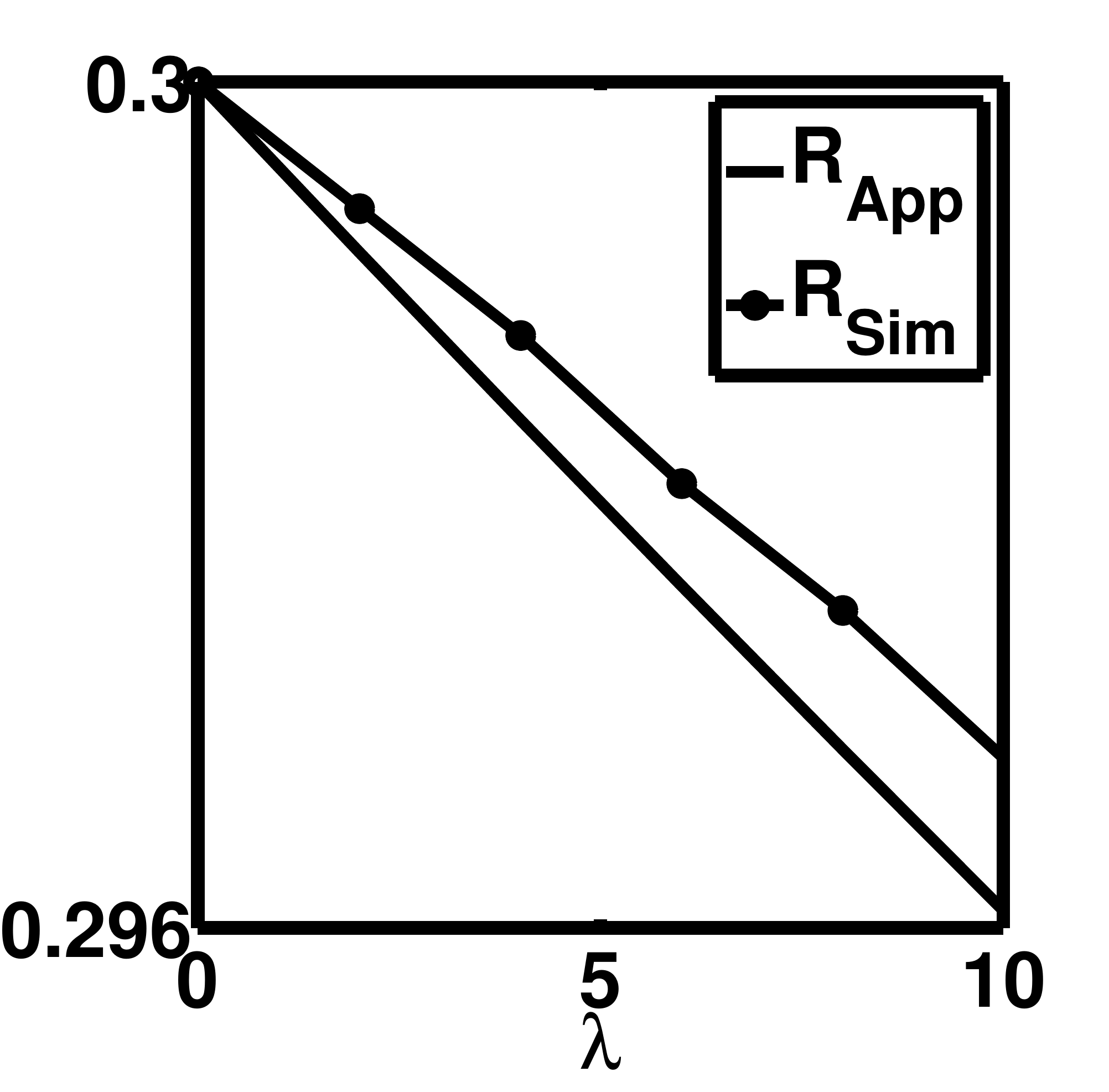}
  \caption{Exponential}
      \label{SQ2 fig:Sce2_exp}
  	         \end{subfigure}       
	           		\begin{subfigure}[b]{.24\textwidth}
  \includegraphics[width=0.99\textwidth]{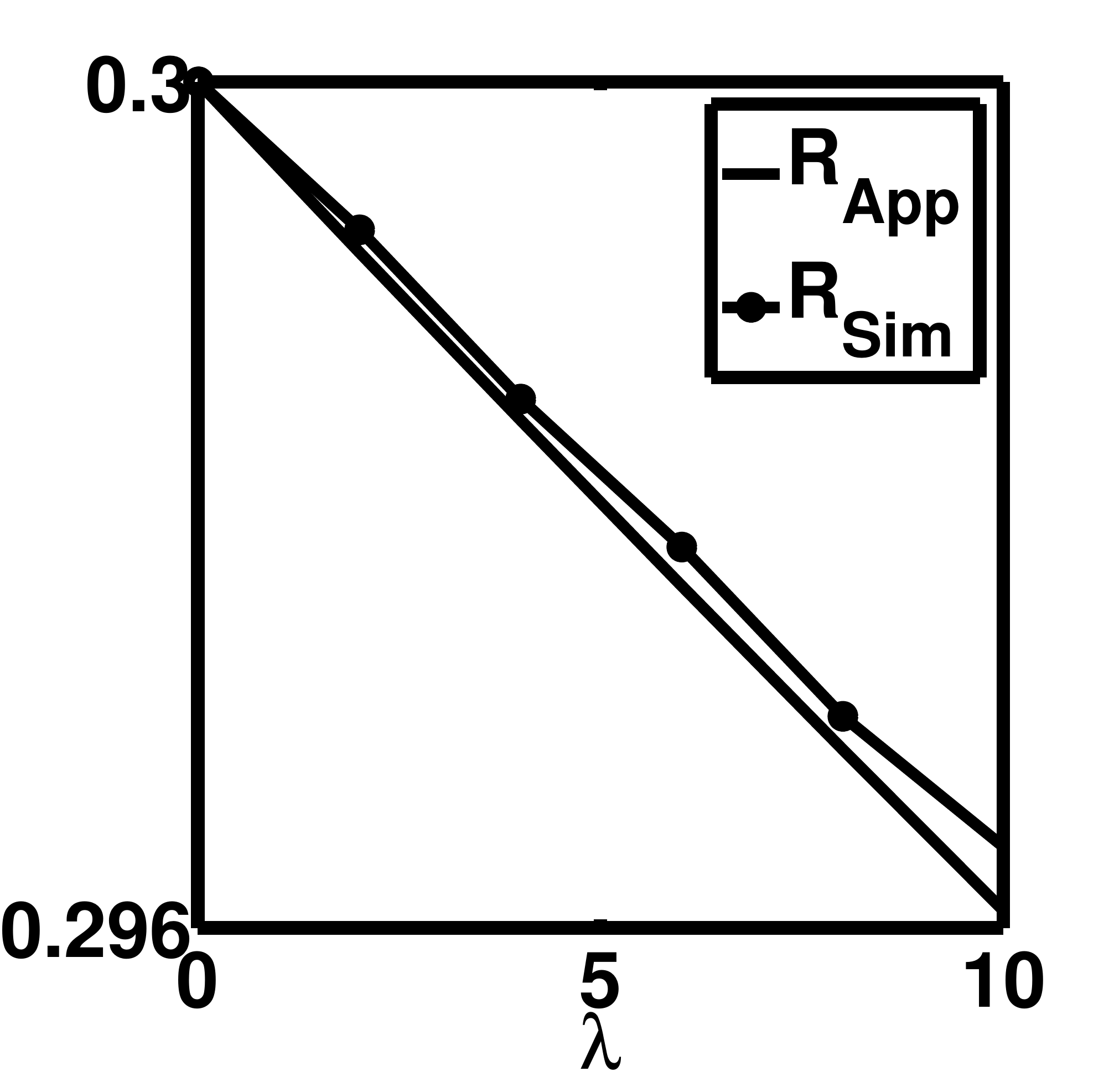}
  \caption{Weibull}
        \label{SQ2 fig:Sce2_weibull}
  	         \end{subfigure}	       
		\begin{subfigure}[b]{.24\textwidth}
  \includegraphics[width=0.99\textwidth]{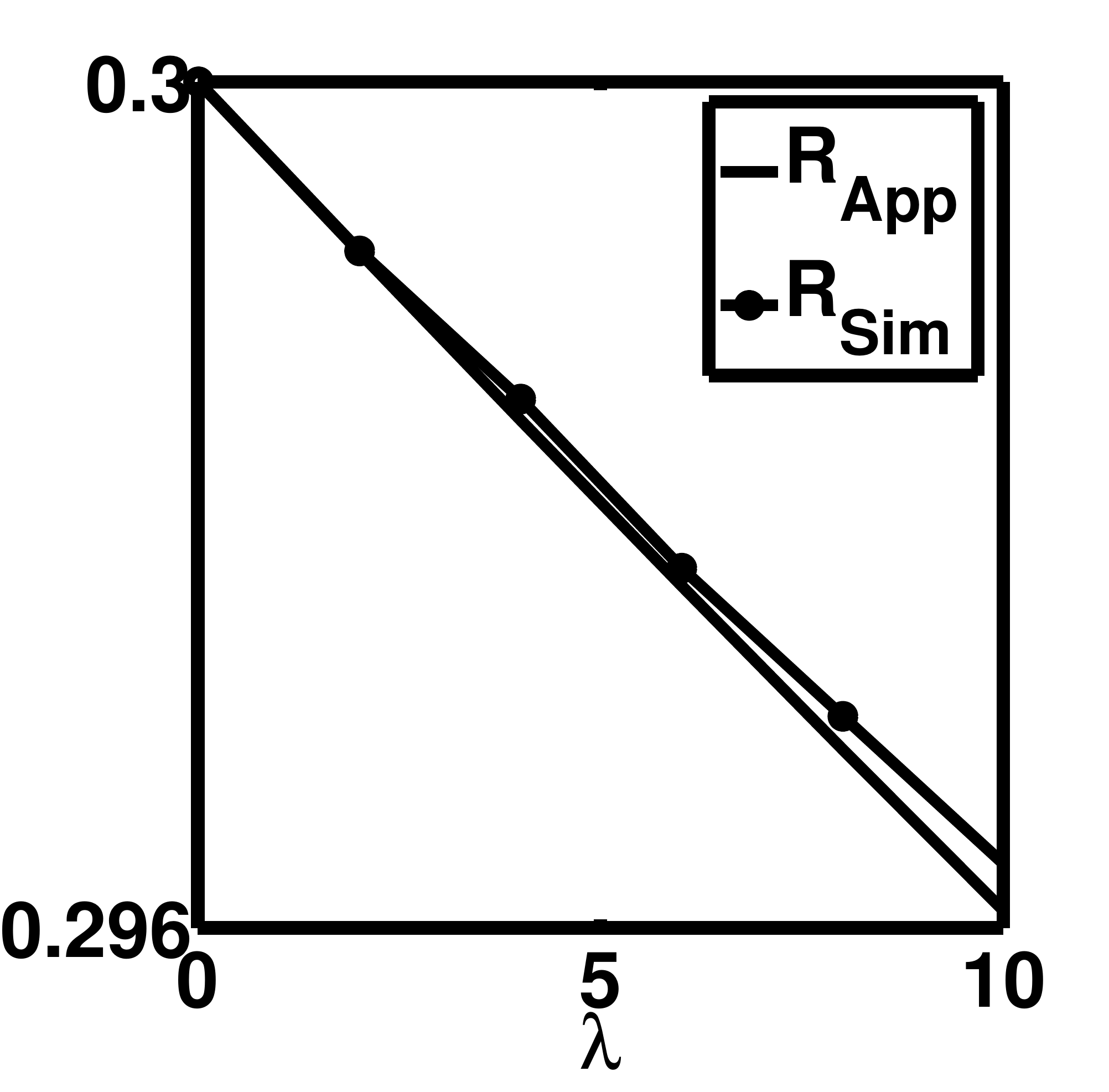}
  \caption{Deterministic}
        \label{SQ2 fig:Sce2_deter}
  	         \end{subfigure}
	         \caption{Scenario 2}
	         \label{SQ2 fig:Sce2}
\end{figure*}

\begin{figure*}
\centering
		\begin{minipage}[b]{.25\textwidth}
  \includegraphics[width=0.99\textwidth]{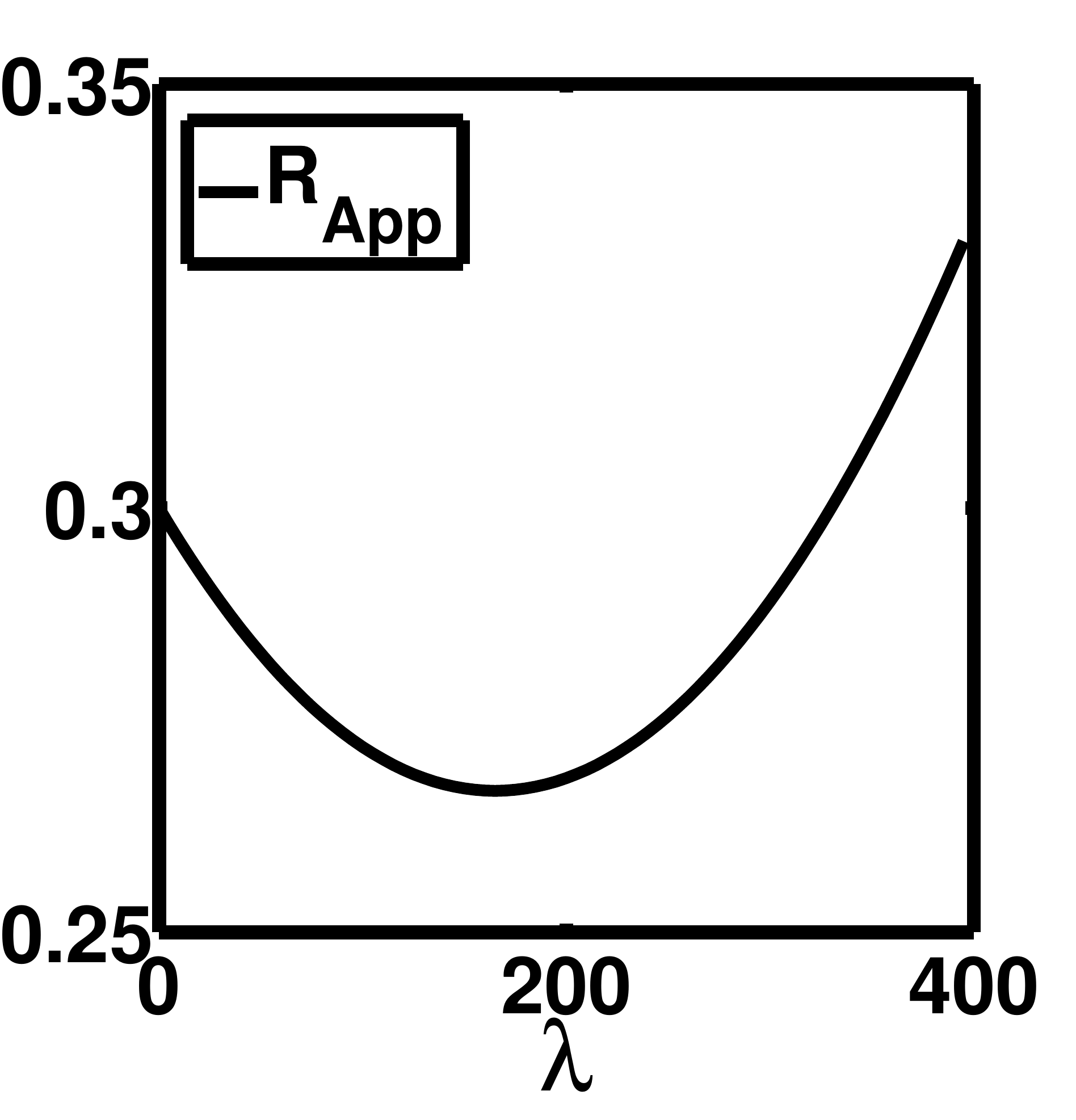}
  \caption{Scenario~2.}
      \label{SQ2 fig:Sce2_app}
  	         \end{minipage}       
\end{figure*}

\begin{figure*}
\centering
\begin{subfigure}[b]{.24\textwidth}
  \includegraphics[width=0.99\textwidth]{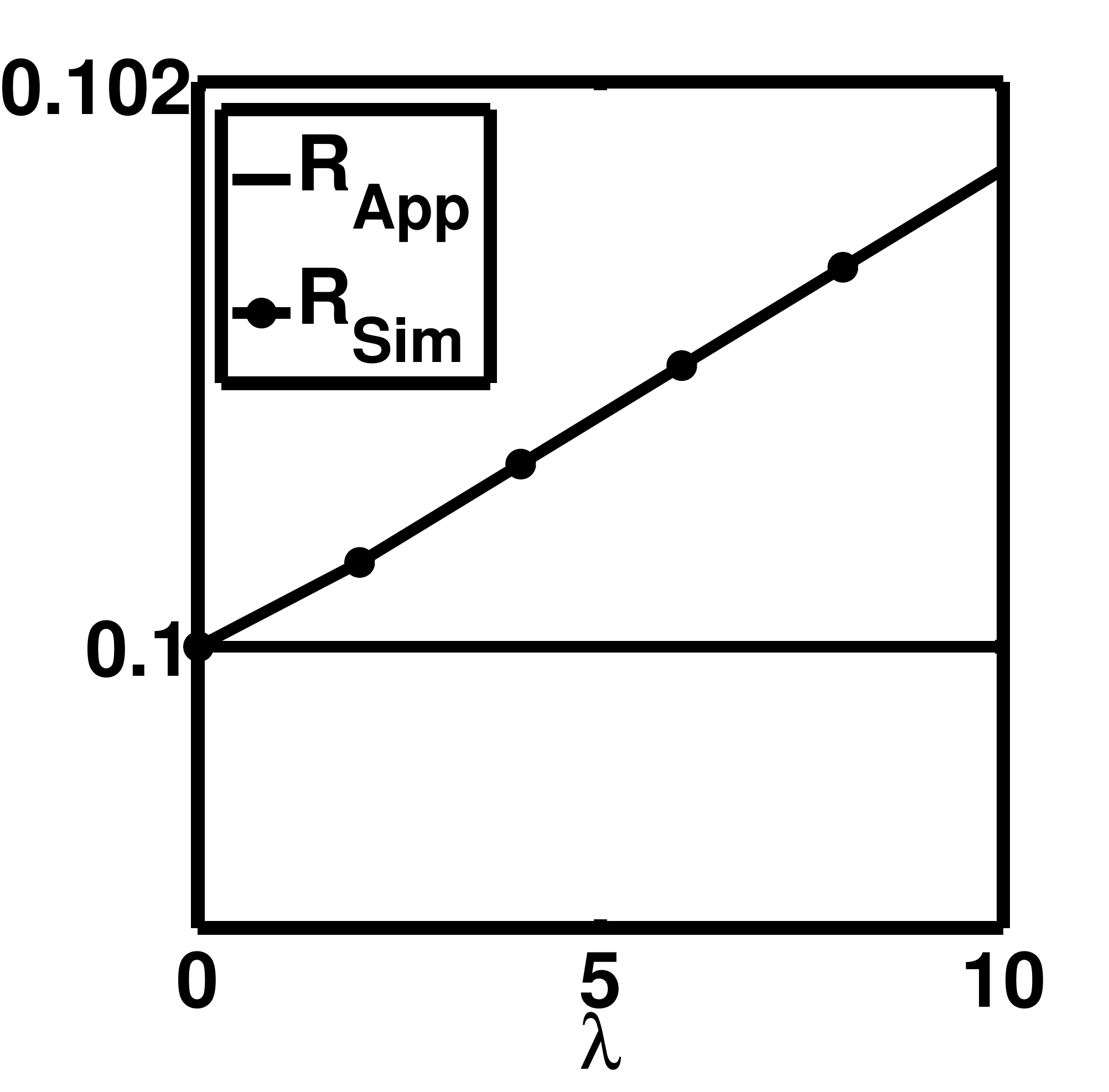}
  \caption{Hyper-exponential}
      \label{SQ2 fig:Sce3_hyper}
  	         \end{subfigure}    
		\begin{subfigure}[b]{.24\textwidth}
  \includegraphics[width=0.99\textwidth]{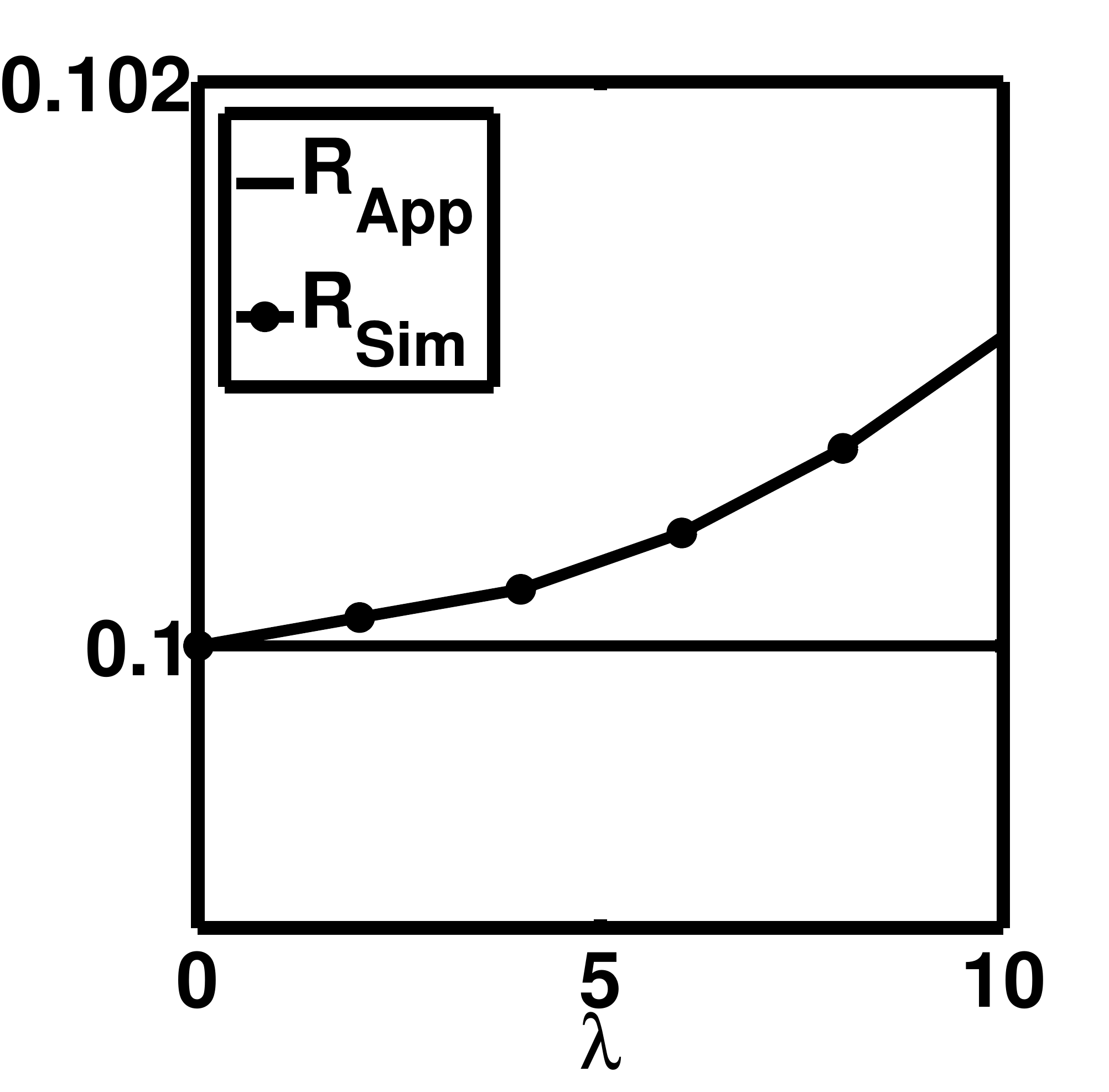}
  \caption{Exponential}
      \label{SQ2 fig:Sce3_exp}
  	         \end{subfigure}       
	           		\begin{subfigure}[b]{.24\textwidth}
  \includegraphics[width=0.99\textwidth]{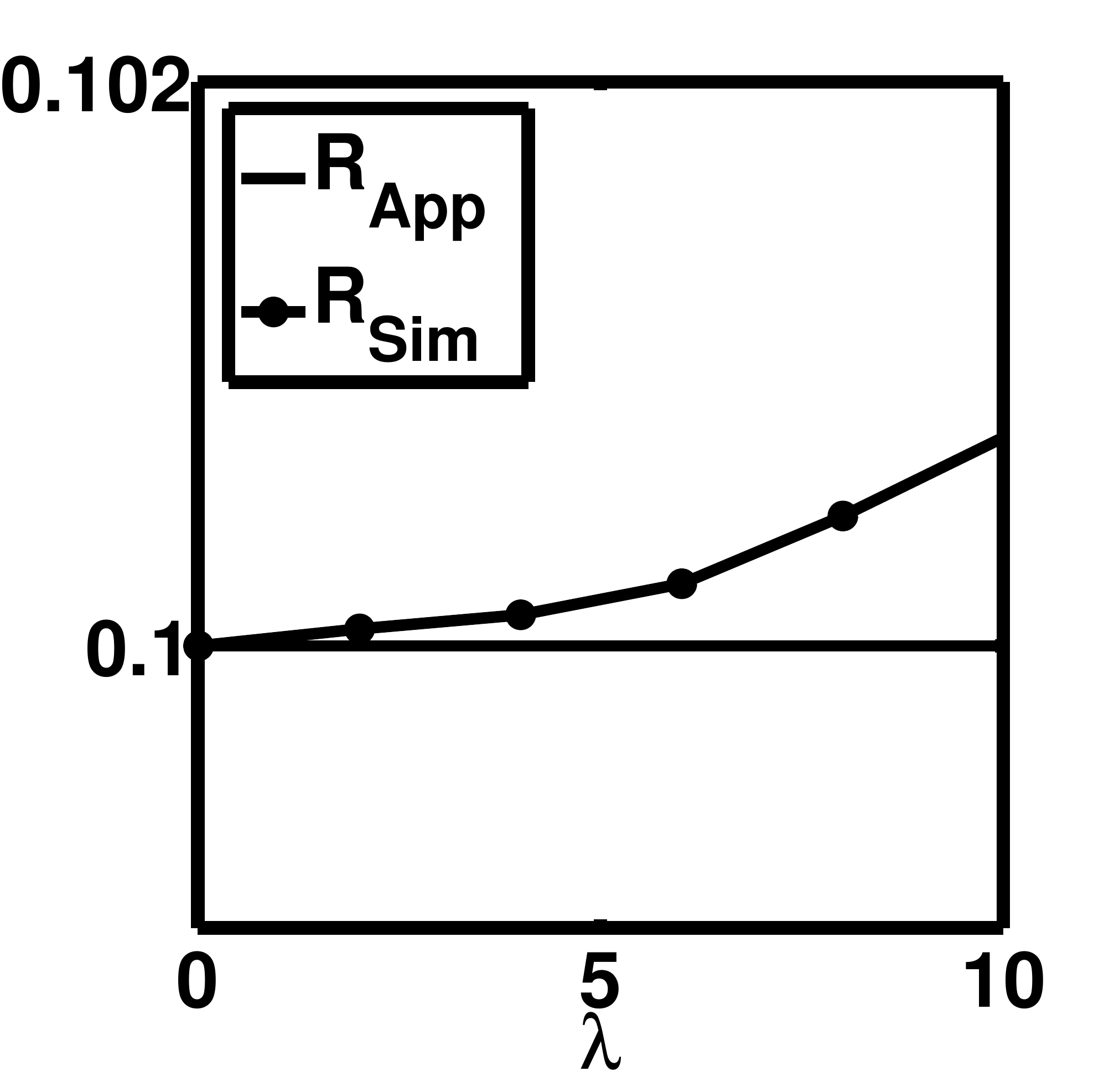}
  \caption{Weibull}
        \label{SQ2 fig:Sce3_weibull}
  	         \end{subfigure}	       
		\begin{subfigure}[b]{.24\textwidth}
  \includegraphics[width=0.99\textwidth]{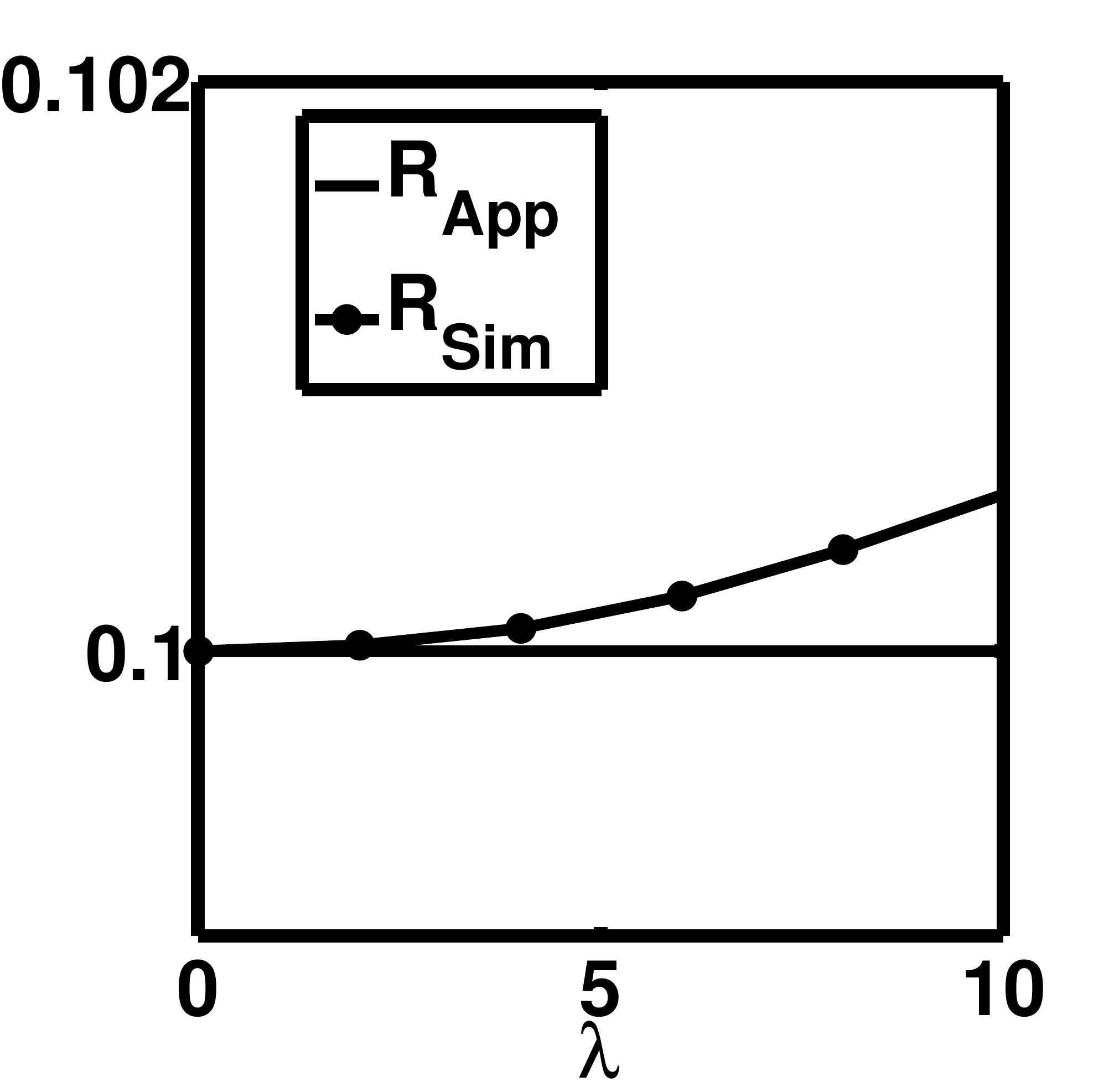}
  \caption{Deterministic}
        \label{SQ2 fig:Sce3_deter}
  	         \end{subfigure}
	         \caption{Scenario 3}
	         \label{SQ2 fig:Sce3}
\end{figure*}

We are interested in the behavior of the average job response time in lightly loaded situations,
and stability is therefore not a concern here as mentioned earlier.
Three different scenarios were explored. 
In Scenarios 1 and 2 there are two types of servers, namely slow servers with capacity $C_{\text{slow}}$ bytes/sec 
and fast servers with capacity $C_{\text{fast}}$ bytes/sec. In Scenario 3 all the servers have the same capacity:

\begin{itemize}
\item Scenario 1: $K=10$ servers. $5$ slow servers with capacity $C_{\text{slow}}=2$ bytes/sec 
and $5$ fast servers with capacity $C_{\text{fast}}=10$ bytes/sec. See Figure~\ref{SQ2 fig:Sce1}. 

\item Scenario 2: $K=100$ servers. $50$ slow servers with capacity $C_{\text{slow}}=2$ bytes/sec 
and $50$ fast servers with capacity $C_{\text{fast}}=10$ bytes/sec. See Figure~\ref{SQ2 fig:Sce2}. 
                                
\item Scenario 3: $K=10$ servers with $C_1 = \ldots = C_{10} = 10$ bytes/sec. See Figure~\ref{SQ2 fig:Sce3}. 

\end{itemize}

In the figures we use
$R_{\rm Sim}$ to denote the average job response time obtained by simulation.
Also  the subscript $p\star$ in the quantities $R_{p\star}(0+)$ and $R_{p\star}^{\prime} (0+) $ refers to Scenario $p$
under distribution $\star$
where $\star$ corresponds to the hyper-exponential~(H), exponential~(E), Weibull~(W) or deterministic~(D) distribution, respectively.

Let ${\rm CV}_{p\star} $ denote the coefficient of variation corresponding to Scenario $p$ under distribution $\star = H, E,W,D$.
Then ${\rm CV}_{iH}=1.4$, ${\rm CV}_{iE}=1$, ${\rm CV}_{iW}=0.52$ and ${\rm CV}_{iD}=~0$ for $i=1,2,3$. 
The simulations do confirm the structural insights  gleaned from the light traffic derivatives for non-homogeneous servers; see Section \ref{sec:Discussion}:
(i) For all distributions, the average job response time decreases
as $\lambda$ increases over a small neighborhood of $\lambda = 0$;
(ii) Over that small interval, performance seems nearly insensitive to
the variability of $\sigma$ (as measured by its coefficient of variation).

Although in Scenario 1 and Scenario 2 there is an equal proportion of slow and fast servers, 
with $R_{p\star}(0+)=0.3000$ for $p=1,2$ and for $\star=H, E, W, D$, 
the impact of the variability in server speeds is seen to diminish with increasing $K$  since
$R^{\prime}_{1\star} (0+)=-0.0044$ and $R^{\prime}_{2\star} (0+)=  -4.0404\cdot 10^{-4}$
for $\star=H, E, W, D$.

Figure \ref{SQ2 fig:Sce2} is a zoom of  Figure \ref{SQ2 fig:Sce2_app} that displays only this common approximation $R_{\rm App}(\lambda)$.
Although in Figure \ref{SQ2 fig:Sce2_app} the response time seems to be a straight line in a small interval of $\lambda$, after a while it also increases.
Since $\bE{\sigma} = 1 $ for all four cases $\star=H,E,W,D$, and the approximation (\ref{SQ2 eq:LT_App_degree2}) that we use depends only on the first moment, Figure \ref{SQ2 fig:Sce2_app} is the same for all distributions considered here.

In Figure \ref{SQ2 fig:Sce3} we observe the aforementioned property for homogeneous servers; 
$R_{\rm App} (\lambda) $ becomes a constant line while the simulation results show that the average response time of a job is increasing.


\section{Review of the light traffic theory \`a la Reiman-Simon}
\label{sec:ReviewLightTrafficTheory}

The light traffic analysis presented here uses an ingenious approach proposed 
by  \citet{RS89} to compute successive light-traffic derivatives in the sense of (\ref{eq:LimitDerivatives}):
Imagine that the system starts at $t=-\infty$, so that its stationary regime will have been reached
at time $t=0$. Enters a {\em tagged} job at time $t=0$ whose expected response time therefore coincides 
with the expected stationary response time.
With this in mind, the $n^{th}$ derivative of the expected stationary response time
at $\lambda = 0+$ (namely (\ref{eq:LimitDerivatives}))
is then shown to be computable in terms of  the expected response time 
of the tagged customer in a {\em scenario}
where {\em exactly} $n$ jobs (other than the tagged job) are allowed into the system.
Details are outlined next.

\subsection{The framework}
\label{subsec:Framework}
  
On the way to describing the Reiman-Simon approach to light traffic 
we find it convenient to introduce the following terminology and notation:
With $t$ in $\mathbb{R}$, a job arriving at time $t$, hereafter referred to as a $t$-job,
has two rvs $\sigma_t$ and $\Sigma_t$ associated with it --
The $\mathbb{R}_+$-valued rv $\sigma_t$ stipulates
the amount of work (in bytes) requested by the $t$-job 
from the system, while the rv $\Sigma_t$ 
is an (unordered) pair of servers from amongst 
the $K$ available servers. 
The $t$-job is assigned to a server $\nu_t$ 
selected in $\Sigma_t$ according to the power-of-two policy 
(with a random tie-breaker).
We shall refer to the rvs $(\sigma_t,\Sigma_t)$ 
as the characteristic pair of the $t$-job.

As expected, we sometimes refer to the $0$-job 
with characteristics $(\sigma_0,\Sigma_0)$ 
as the {\em tagged} job.
The Reiman-Simon approach to light traffic
focuses on the performance of this tagged job 
under scenarios of increasing complexity. To define them,
fix $n=0,1, \ldots $. 
Interpret every $n$-uple $(t_1, \ldots , t_n)$ 
in $\mathbb{R}^n$ as the arrival epochs
of $n$ jobs into the system. 
For each $i=1, \ldots , n$,  
we lighten the notation by denoting the characteristic
pair $(\sigma_{t_i}, \Sigma_{t_i})$
of the $t_i$-job arriving at time $t_i$ simply by
$(\sigma_i, \Sigma_i)$.
Throughout the following conditions are assumed to be enforced:

\begin{enumerate}
\item The rvs $\{ \sigma_0, \sigma_{1}, \ldots , \sigma_{n} \}$ are i.i.d. $\mathbb{R}_+$-valued rvs, 
each distributed according to 
the probability distribution $F$, namely
\[
\bP{ \sigma_{i} \leq x } = F(x),
\quad 
\begin{array}{c}
x \geq 0 \\
i=0,1, \ldots , n. \\
\end{array}
\]

\item The rvs $\{ \Sigma_0, \Sigma_1, \ldots , \Sigma_n \}$ are i.i.d.
${\cal P}_2(K)$-valued rvs, each of which is uniformly distributed on ${\cal P}_2(K) $ with
\[
\bP{ \Sigma_i = T  } = \frac{1}{ {K \choose 2 }},
\quad 
\begin{array}{c}
T \in {\cal P}_2(K) \\
i=0,1, \ldots , n. \\
\end{array}
\]

\item The collections of rvs  $\{ \sigma_0, \sigma_1, \ldots , \sigma_n \}$ and $\{ \Sigma_0, \Sigma_1, \ldots , \Sigma_n \}$ 
are mutually independent
\end{enumerate}

We shall also have use for the rvs $\nu^\star_0, \nu^\star_1, \ldots , \nu^\star_n$ associated with 
the random pairs $\Sigma_0, \Sigma_1, \ldots , \Sigma_n $, and defined in the following manner:
For each $i=0,1, \ldots , n$, {\em conditionally} on $\Sigma_i$, 
the rv $\nu^\star_i$ is an $\Sigma_i$-valued rv which is uniformly
distributed on $\Sigma_i$ -- We shall write
\[
[ \nu^\star_i | \Sigma_i ] \sim {\cal U}(\Sigma_i).
\]
It is always understood 
that the rvs $\nu^\star_0, \nu^\star_1, \ldots , \nu^\star_n$ 
are conditionally mutually independent given the $2(n+1)$ rvs
$\sigma_0, \sigma_1, \ldots , \sigma_n ,  
\Sigma_0, \Sigma_1, \ldots , \Sigma_n $ with
\[
[ \nu^\star_i | \sigma_0, \ldots , \sigma_n, \Sigma_0 , \ldots , \Sigma_n ] \sim {\cal U}(\Sigma_i),.
\quad i=0,1, \ldots , n.
\]
Under the enforced assumptions, we readily conclude that
the rvs $\nu^\star_0, \nu^\star_1, \ldots , \nu^\star_n$ are i.i.d. rvs, 
each of which is uniformly distributed on
$\{ 1, \ldots , K \}$ (as shown in Proposition \ref{prop:ExpectedR_0}).

\subsection{Computing the derivatives}

Fix $n=1,2, \ldots $. For each $(t_1, \ldots , t_n)$ in $\mathbb{R}^n$, let the rv $R_n(t_1, \ldots , t_n)$ 
denote the response time of the tagged job under the scenario that  {\em in addition}
to the tagged job, only $n$ jobs are allowed to enter the system 
over $\mathbb{R}$, say at times $t_1, \ldots , t_n$, with characteristic pairs
$(\sigma_1, \Sigma_1), \ldots , (\sigma_n,\Sigma_n)$ as
defined earlier. Note that $R_n(t_1, \ldots , t_n)$  depends on the rvs
$\{ \sigma_0, \sigma_1, \ldots , \sigma_n \}$, 
$\{ \Sigma_0, \Sigma_1, \ldots , \Sigma_n \}$ 
and
$\{ \nu^\star_0, \nu^\star_1, \ldots , \nu^\star_n \}$
in a complicated manner through the scheduling policy used.
We shall write
\begin{equation}
\widehat R_n(t_1, \ldots , t_n )
= \bE{ R_n(t_1, \ldots , t_n ) }.
\label{eq:R_n}
\end{equation}

Under some appropriate integrability conditions, Reiman and Simon show that the 
light-traffic derivatives in the sense of (\ref{eq:LimitDerivatives})
can be expressed in terms of the quantities  (\ref{eq:R_n}) -- Here we consider the cases $n=0,1,2$:
Using Theorems 1 and 2  in \citep[pp. 29-30]{RS89} for $n=0,1,2$ we collect the expressions
\begin{equation}
R (0+)
=
\lim_{\lambda \downarrow 0} R (\lambda )
= \widehat R_0
\label{eq:ExpressionFor_n=0}
\end{equation}
with $\widehat R_0$ defined in Section \ref{sec:n=0},
\begin{equation}
R^{\prime} (0+)
=
\lim_{\lambda \downarrow 0} 
\frac{dR}{d\lambda} (\lambda )
= 
\int_{\mathbb{R}} 
\left ( \widehat R_1(t) - \widehat R_0 \right ) dt
\label{eq:ExpressionFor_n=1}
\end{equation}
and
\begin{equation}
R^{\prime\prime} (0+)
= 
\lim_{\lambda \downarrow 0} 
\frac{d^2R}{d\lambda^2} (\lambda )
=
\int_{\mathbb{R}}
\left (  \int_{\mathbb{R}} 
( \widehat R_2(s,t) -  \widehat R_1(s) - \widehat R_1(t) +  \widehat R_0 ) dt
 \right ) ds
\label{eq:ExpressionFor_n=2}
\end{equation}

\section{The case $n=0$}
\label{sec:n=0}

The case $n=0$ is slightly different and corresponds to the scenario when besides the tagged customer,
no other job enters  over the entire horizon $(-\infty, \infty)$.
Let $R_0$ denote the response time of the tagged job 
under these circumstances. Obviously,
under the power-of-two scheduling strategy, we have
\begin{equation}
R_0 = \frac{ \sigma_0}{C_{\nu_0}} \quad \mbox{with $\nu_0 = \nu^\star_0$}
\label{eq:n=0}
\end{equation}
because  in the absence of any
other job in the system, the tagged job is necessarily
assigned to server $\nu^\star_0$.
Somewhat in analogy with earlier notation we write
\[
\widehat R_0 = \bE{ R_0 }.
\]

\begin{proposition}
{\sl Under the enforced assumptions, 
the rv $\nu^\star_0$ is uniformly distributed over
$\{1, \ldots , K \}$ with
\begin{equation}
\bP{ \nu^\star_0 = k } 
= \frac{1}{K},
\quad k=1, \ldots , K
\label{eq:UniformPMF}
\end{equation}
and the relation
\begin{equation}
\widehat R_0  
= 
\left ( \frac{1}{K} \sum_{k=1}^K \frac{1}{C_k} \right ) \cdot \bE{ \sigma}
\label{eq:ExpectedR_0}
\end{equation}
holds.
}
\label{prop:ExpectedR_0}
\end{proposition}

With $\Gamma$ given at (\ref{eq:Gamma}) it
will often be convenient to write
(\ref{eq:ExpectedR_0}) more compactly as
\begin{equation}
\widehat R_0  
=
\frac{ \Gamma}{K} \cdot \bE{ \sigma }.
\label{eq:ExpectedR_0Alternate}
\end{equation}

\myproof
For each $k=1, \ldots , K$, the definition of $\nu_0^\star$ gives
\begin{eqnarray}
\bP{ \nu^\star_0 = k }
&=&
\sum_{\ell=1, \ \ell \neq k }^K 
\bP{ \Sigma_0 = \{ k, \ell \}, \nu^\star_0 = k }
\nonumber \\
&=&
\sum_{\ell=1, \ \ell \neq k }^K 
\bP{ \nu^\star_0 = k | \Sigma_0 = \{ k, \ell \} }
\bP{ \Sigma_0 = \{ k, \ell \} }
\nonumber \\
&=& 
(K-1) \cdot \frac{1}{2} \cdot \frac{2}{K(K-1)} 
= \frac{1}{K}.
\end{eqnarray}

As pointed earlier, we necessarily have $\nu_0 = \nu^\star_0$. 
The rvs $\nu^\star_0$ and $\sigma_0$ being independent,
we then obtain from (\ref{eq:n=0}) that
\[
\widehat R_0
= 
\bE{ \frac{\sigma_0}{C_{\nu^\star_0}} }
= 
\bE{ \sigma_0 }
\cdot \bE{ \frac{1}{C_{\nu^\star_0}} },
\]
and the conclusion (\ref{eq:ExpectedR_0})
readily follows from (\ref{eq:UniformPMF}).
\myendpf

According to the Reiman-Simon theory, we have $R(0+) = \widehat R_0$
and Proposition \ref{prop:ExpectedR_k=0} is established with the help of
(\ref{eq:ExpectedR_0}).

\section{An auxiliary result}
\label{sec:AuxiliaryResult}

The cases $n=1$ and $n=2$ are computationally more involved. 
The technical result discussed next will
simplify the presentation by isolating an evaluation which is
repeatedly carried out during the analysis.  
This auxiliary result is given in a setting that mimics power-of-two
scheduling with only two customers present:

Fix $y < 0$.
In addition to the tagged job arriving at time $t=0$ 
with characteristic pair $(\sigma_0, \Sigma_0)$, assume that
another job arrives at time $y$ with (random) service requirement $\tau$.
This $y$-job is then assigned to the server $\gamma$,
with $\gamma$ being some $\{1, \ldots , K \}$-valued rv,
while the tagged job is assigned to the server $\gamma_0$ (in $\Sigma_0$) 
in accordance with the power-of-two scheduling policy.
Thus, if $y + \frac{\tau}{C_{\gamma}} \leq 0$, then
$\gamma_0 = \nu^\star_0$. On the other hand, if
$y + \frac{\tau}{C_{\gamma}} > 0$, then 
the operational rules of the power-of-two scheduling policy will preclude
the tagged job to be assigned to server $\gamma$:
Indeed, if $\gamma$ is not in $\Sigma_0$, then $\gamma_0 = \nu^\star_0$ again, 
while if $\gamma$ is an element of $\Sigma_0$, then $\gamma_0$ 
is necessarily the other server in the pair $\Sigma_0$, i.e.,
the one different from $\gamma$.  
In this scenario  $\gamma$ is a rv given {\em a priori}, and should be 
thought as a place holder for a server assignment rv determined via
power-of-two scheduling under various circumstances. On the other hand,
$\gamma_0$ depends on $y$, $\tau$, $\gamma$ and $\Sigma_0$ (as well as $\nu^\star_0$).
The explicit dependence on these quantities will be dropped from the notation.

For reasons that will become apparent in subsequent developments,
we also introduce an event $E$ (to be specified later).

\begin{lemma}
{\sl Given are the rvs $\1{E}$, $\tau$, $\gamma$, $\sigma_0$, $\Sigma_0$
and $\nu^\star_0$. 
We assume that 
(i) the rv $\nu^\star_0$ is uniformly distributed on $\Sigma_0$
conditionally 
on all the other rvs $\1{E}$, $\tau$, $\gamma$, $\sigma_0$ and $\Sigma_0$;
(ii) the collections of rvs $\{ \1{E}, \tau, \gamma \}$  and $\{ \nu^\star_0 , \Sigma_0 , \sigma_0\}$
are independent; 
and (iii) the rvs $\Sigma_0$ and $\sigma_0$ are independent.
Then, for each $y<0$ and each $k=1, \ldots , K$, we have
\begin{eqnarray}
\bE{ \1{E} \1{\gamma = k} \frac{\sigma_0}{C_{\gamma_0}} }
&=&
\bP{ E, \gamma = k, y + \frac{\tau}{C_k} \leq 0 }
\cdot
\widehat R_0
\nonumber \\
& &
~+ \frac{1}{K-1}
\bP{ E, \gamma = k, y + \frac{\tau}{C_k} > 0 } 
\left ( \Gamma - \frac{1}{C_k} \right ) \cdot \bE{\sigma_0} 
\label{eq:Auxiliary}
\end{eqnarray}
with $\gamma_0$ as defined earlier.
\label{lem:Auxiliary}
} 
\end{lemma}

Recall that under the enforced assumptions, the rv $\sigma_0$ is
independent of the collection of rvs $\{ \nu^\star_0 , \Sigma_0 \}$; 
see Section \ref{subsec:Framework}.

\myproof
Fix $k=1, \ldots , K $.
We start with the natural decomposition
\begin{eqnarray}
\lefteqn{
\bE{ \1{E} \1{\gamma = k} \frac{\sigma_0}{C_{\gamma_0}} }
} & &
\label{eq:Auxiliary1} \\
&=&
\bE{ \1{E} \1{\gamma = k}\1{y + \frac{\tau}{C_{\gamma}} \leq 0} 
\frac{\sigma_0}{C_{\gamma_0}} }
+
\bE{ \1{E} \1{\gamma = k}\1{y + \frac{\tau}{C_{\gamma}} > 0 }
\frac{\sigma_0}{C_{\gamma_0}} } .
\nonumber
\end{eqnarray}

For the first term, the definition of $\gamma_0$ leads to
\begin{eqnarray}
\bE{ \1{E} \1{\gamma = k}\1{y + \frac{\tau}{C_{\gamma}} \leq 0}
\frac{\sigma_0}{C_{\gamma_0}} }
&=&
\bE{ \1{E} \1{\gamma = k}\1{y + \frac{\tau}{C_k} \leq 0}
\frac{\sigma_0}{C_{\nu^\star_0}} }
\nonumber \\
&=&
\bE{ \1{E} \1{\gamma = k}\1{y + \frac{\tau}{C_k} \leq 0} }
\bE{ \frac{\sigma_0}{C_{\nu^\star_0}} } 
\nonumber \\
&=&
\bP{ E, \gamma = k, y + \frac{\tau}{C_k} \leq 0 }
\cdot
\widehat R_0
\label{eq:Auxiliary2}
\end{eqnarray}
since the collections $\{ \1{E}, \gamma, \tau \}$ and $\{ \nu^\star_0, \sigma_0 \}$
are independent under the enforced assumptions.

We further decompose the second term in (\ref{eq:Auxiliary1}) to obtain
\begin{eqnarray}
\lefteqn{
\bE{ \1{E} \1{\gamma = k}\1{y + \frac{\tau}{C_{\gamma}} > 0}
\frac{\sigma_0}{C_{\gamma_0}} }
} & &
\nonumber \\
&=&
\bE{ \1{E} \1{\gamma = k}\1{y+ \frac{\tau}{C_k} > 0}\1{ k \notin \Sigma_0}
\frac{\sigma_0}{C_{\nu^\star_0}} }
\nonumber \\
& &
~+
\bE{ \1{E} \1{\gamma = k}\1{y + \frac{\tau}{C_k} > 0}\1{ k \in \Sigma_0}
\frac{\sigma_0}{C_{\gamma_0}} }.
\label{eq:Auxiliary3}
\end{eqnarray}
It is plain that
\begin{eqnarray}
\lefteqn{
\bE{ \1{E} \1{\gamma = k}\1{y+ \frac{\tau}{C_k} > 0}\1{ k \in \Sigma_0}
\frac{\sigma_0}{C_{\gamma_0}} }
} & &
\nonumber \\
&=&
\sum_{\ell=1, \ell \neq k }^K
\bE{ \1{E} \1{\gamma = k}\1{y + \frac{\tau}{C_k} > 0}
\1{ \Sigma_0 = \{ k, \ell \} }
\frac{\sigma_0}{C_{\gamma_0}} }
\nonumber \\
&=&
\sum_{\ell=1, \ell \neq k }^K
\bE{ \1{E} \1{\gamma = k}\1{y + \frac{\tau}{C_k} > 0}
\1{ \Sigma_0 = \{ k, \ell \} }
\frac{\sigma_0}{C_\ell} }
\nonumber \\
&=&
\frac{2}{K(K-1)}
\sum_{\ell=1, \ell \neq k }^K
\bP{ E, \gamma = k, y + \frac{\tau}{C_k} > 0 }
\frac{\bE{\sigma_0}}{C_\ell} 
\nonumber \\
&=&
\frac{2}{K(K-1)}
\left (
\sum_{\ell=1, \ell \neq k }^K
\frac{\bE{\sigma_0}}{C_\ell} 
\right )
\bP{ E, \gamma = k, y + \frac{\tau}{C_k} > 0 }
\nonumber \\
&=&
\frac{2}{K(K-1)}
\left ( \Gamma - \frac{1}{C_k} \right )
\bP{ E, \gamma = k, y + \frac{\tau}{C_k} > 0 }
\cdot \bE{ \sigma_0}
\label{eq:Auxiliary4}
\end{eqnarray}
since $\gamma_0=\ell$ if $\Sigma_0 = \{ k, \ell \}$ 
when $\gamma = k$ and $ y + \frac{\tau}{C_k} > 0 $.

On the other hand, the definition of $\nu^\star_0$ implies
\begin{eqnarray}
\lefteqn{
\bE{ \1{E} \1{\gamma = k}\1{y + \frac{\tau}{C_k} > 0}\1{ k \notin \Sigma_0}
\frac{\sigma_0}{C_{\nu^\star_0}} }
} & &
\nonumber \\
&=&
\sum_{  T \in {\cal P}_2(K) }
\bE{ \1{E} \1{\gamma = k}\1{y + \frac{\tau}{C_k} > 0}\1{ k \notin \Sigma_0}
\1{ \Sigma_0 = T }
\frac{\sigma_0}{C_{\nu^\star_0}} }
\nonumber \\
&=&
\sum_{a=1, a \neq k}^K
\left (\sum_{b=1, b \neq k}^{a-1}
\bE{ \1{E} \1{\gamma = k}\1{y + \frac{\tau}{C_k} > 0}
\1{ \Sigma_0 = \{a,b\} }
\frac{\sigma_0}{C_{\nu^\star_0}} }
\right )
\nonumber \\
&=&
\sum_{a=1, a \neq k}^K
\left (\sum_{b=1, b \neq k}^{a-1}
\bP{ E, \gamma = k, y + \frac{\tau}{C_k} > 0 }
\bE{ \1{ \Sigma_0 = \{a,b\} } \frac{\sigma_0}{C_{\nu^\star_0}} }
\right )
\nonumber \\
&=&
\sum_{a=1, a \neq k}^K
\bP{ E, \gamma = k, y + \frac{\tau}{C_k} > 0 }
\left (\sum_{b=1, b \neq k}^{a-1}
\bP{ \Sigma_0 = \{a,b\} } 
\cdot \frac{1}{2}
\left ( \frac{1}{C_a} + \frac{1}{C_b} \right )
\right )
\cdot \bE{\sigma_0}
\nonumber \\
&=&
\frac{1}{K(K-1)}
\sum_{a=1, a \neq k}^K
\bP{ E, \gamma = k, y + \frac{\tau}{C_k} > 0 }
\left (\sum_{b=1, b \neq k}^{a-1}
\left ( \frac{1}{C_a} + \frac{1}{C_b} \right )
\right )
\cdot \bE{\sigma_0} 
\nonumber \\
&=&
\frac{1}{K(K-1)}
\left (
\sum_{a=1, a \neq k}^K
\left (\sum_{b=1, b \neq k}^{a-1}
\left ( \frac{1}{C_a} + \frac{1}{C_b} \right )
\right )
\right )
\bP{ E, \gamma = k, y + \frac{\tau}{C_k} > 0 }
\cdot \bE{\sigma_0} .
\label{eq:Auxiliary6}
\end{eqnarray}
In Appendix B we show that
\begin{equation}
\sum_{ a=1 , a \neq k }^K
\left (\sum_{b=1, b \neq k}^{a-1}
\left ( \frac{1}{C_a} + \frac{1}{C_b}  \right ) 
\right )
= (K-2) \left ( \Gamma - \frac{1}{C_k} \right ),
\label{eq:Sigma_k}
\end{equation}
so that (\ref{eq:Auxiliary6}) can be written more compactly as
\begin{eqnarray}
\lefteqn{
\bE{ \1{E}  \1{\gamma = k}\1{y + \frac{\tau}{C_k} > 0}\1{ k \notin \Sigma_0}
\frac{\sigma_0}{C_{\nu^\star_0}} }
} & &
\nonumber \\
&=&
\frac{K-2}{K(K-1)} 
\left ( \Gamma - \frac{1}{C_k} \right )
\bP{E, \gamma = k, y + \frac{\tau}{C_k} > 0 }
\cdot \bE{\sigma_0} .
\label{eq:Auxiliary5}
\end{eqnarray}

To conclude the proof, substitute 
(\ref{eq:Auxiliary4}) and (\ref{eq:Auxiliary5})
into (\ref{eq:Auxiliary3}).
It yields
\[
\bE{ \1{E} \1{\gamma = k}\1{y + \frac{\tau}{C_\gamma} > 0}
\frac{\sigma_0}{C_{\gamma_0}} }
= 
\frac{1}{K-1} \left ( \Gamma - \frac{1}{C_k} \right )
\bP{ E, \gamma = k, y + \frac{\tau}{C_k} > 0 }
\cdot \bE{\sigma_0},
\]
and combining this last expression with (\ref{eq:Auxiliary2}) we get
the desired result (\ref{eq:Auxiliary})
with the help of (\ref{eq:Auxiliary1}).
\myendpf

\section{The case $n=1$}
\label{sec:n=1}

The analysis of the first derivative is associated with the following scenario:
The tagged job arrives at time $t=0$ with characteristic
pair $(\sigma_0, \Sigma_0)$.
With $t$ in $\mathbb{R}$, in addition to the tagged job,
a single job arrives during the entire horizon $(-\infty, \infty)$,
say at time $t$ with characteristic pair $(\sigma_t, \Sigma_t)$.
The tagged job and this $t$-job are assigned to the 
servers $\nu_0$ (in $\Sigma_0$) and $\nu_t$ (in $\Sigma_t$), 
respectively, in accordance with the power-of-two
scheduling policy.

\subsection{Evaluating $\widehat R_1(t)$}

For each $t$ in $\mathbb{R}$, in accordance with (\ref{eq:R_n}) we have
\begin{equation}
\widehat R_1(t) = \bE{ R_1(t) }
\quad
\mbox{with~} 
R_1(t) = \frac{ \sigma_0 }{ C_{\nu_0} } .
\label{eq:n=1}
\end{equation}
However, with the presence of the $t$-job,
$\nu_0$ does not always coincide
with $\nu^\star_0$, as the determination of $\nu_0$ 
may be affected by whether the
$t$-job completed service
at the time the tagged job arrives.

First some notation:
With $t$ arbitrary in $\mathbb{R}$, set
\begin{equation}
H_k(t)
=
\bP{ C_k t + \sigma_t \leq 0 } \cdot \widehat R_0 
+ \frac{1}{K-1} \left ( \Gamma - \frac{1}{C_k} \right )
\bP{ C_k t + \sigma_t  > 0 } \cdot \bE{\sigma_0}
\label{eq:H_k(t)}
\end{equation}
for each $k=1, \ldots , K$. Note that
\begin{eqnarray}
H_k(t)
&=&
\widehat R_0 \cdot 
\left ( 1 - \bP{ C_k t + \sigma_t > 0 } \right )
+ \frac{1}{K-1} \left ( \Gamma - \frac{1}{C_k} \right ) 
\bE{\sigma_0}
\cdot \bP{ C_k t + \sigma_t > 0 }
\nonumber \\
&=&
\widehat R_0
+
\left ( \frac{1}{K-1} 
\left ( \Gamma - \frac{1}{C_k} \right ) - \frac{\Gamma}{K}  \right ) \bE{\sigma_0}
\cdot \bP{ C_k t + \sigma_t > 0 }
\nonumber \\
&=&
\widehat R_0
+
\frac{1}{K-1} \left ( \frac{\Gamma}{K} - \frac{1}{C_k}  \right )
\bP{ C_k t + \sigma_t > 0 } \cdot \bE{\sigma_0}
\label{eq:H_k(t)_Alternate}
\end{eqnarray}
as we make use of the expression (\ref{eq:ExpectedR_0Alternate}).

\begin{proposition}
{\sl Under the enforced independence assumptions,
we have $\widehat R_1(t) = \widehat R_0$ if $t > 0$, while for $t < 0$ it holds that
\begin{eqnarray}
\widehat R_1(t)
=
\frac{1}{K} \sum_{k=1}^K H_k(t).
\label{eq:ExpectedR_1B}
\end{eqnarray}
}
\label{prop:ExpectedR_1}
\end{proposition}

\myproof
Fix $t$ in $\mathbb{R}$. As we seek to evaluate
$\widehat R_1(t)$ as given by (\ref{eq:n=1}),
two cases  need to be examined: 
If $ t > 0$, then $\nu_0 = \nu^\star_0$, whence
$R_1(t) = R_0$, and the conclusion $\widehat R_1(t) = \widehat R_0$ follows.

If $t < 0$, 
then $\nu_t  = \nu^\star_t$ and we are in the setting
of Lemma \ref{lem:Auxiliary}
with $y=t$, $E=\Omega$, $\tau = \sigma_t$ and $\gamma = \nu^\star_t$ (so that $\gamma_0=\nu_0$):
For each $k=1, \ldots  , K$, the expression
(\ref{eq:Auxiliary}) becomes
\begin{eqnarray}
& & \bE{ \1{\nu^\star_t  = k} \frac{\sigma_0}{C_{\nu_0}} }
\nonumber \\
&=&
\bP{ \nu^\star_t  = k, t + \frac{\sigma_t}{C_k} \leq 0 }
\cdot
\widehat R_0
+ \frac{1}{K-1}
\bP{ \nu^\star_t = k, t + \frac{\sigma_t}{C_k} > 0 }
\left ( \Gamma - \frac{1}{C_k} \right ) \cdot \bE{\sigma_0} 
\nonumber \\
&=&
\frac{1}{K} \cdot H_k(t)
\end{eqnarray}
since the rv $\nu^\star_t$ is independent of $\sigma_t$ 
and uniformly distributed on $\{1, \ldots , K \}$
(as pointed out in Proposition \ref{prop:ExpectedR_0}).
The desired result (\ref{eq:ExpectedR_1B}) now
follows from (\ref{eq:n=1}) upon noting the decomposition
\[
\widehat R_1(t) = \sum_{k=1}^K \bE{ \1{\nu^\star_t  = k} \frac{\sigma_0}{C_{\nu_0}} } .
\]
\myendpf

\subsection{A proof of Proposition \ref{prop:ExpectedR_k=1}}

We can now complete the proof of Proposition \ref{prop:ExpectedR_k=1}:
The expression (\ref{eq:ExpressionFor_n=1}) now takes the form
\begin{eqnarray}
R^{\prime} (0+)
=
\int_{\mathbb{R}} 
\left ( \widehat R_1(t) - \widehat R_0 \right ) dt
=
\int_{-\infty}^0
\left ( \widehat R_1(t) - \widehat R_0 \right ) dt
\label{eq:IntegralFor_n=1}
\end{eqnarray}
as we recall that $\widehat R_1(t) = \widehat R_0$ for $t > 0$.
Next, for $t < 0$, with the help of (\ref{eq:H_k(t)_Alternate}) and (\ref{eq:ExpectedR_1B})
we can rewrite the integrand as
\begin{eqnarray}
\widehat R_1(t) - \widehat R_0
=
 \frac{1}{K}
\sum_{k=1}^K 
\frac{1}{K-1} \left ( \frac{\Gamma}{K} - \frac{1}{C_k}  \right )
\bP{ C_k t + \sigma > 0 } \cdot \bE{\sigma}
\label{eq:Integrand}
\end{eqnarray}
as we recall that
the rvs $\sigma_t$ and $\sigma_0$ are both distributed like $\sigma$.

Inserting this expression back into (\ref{eq:IntegralFor_n=1}) we get
\begin{equation}
\int_{-\infty}^0
\left ( \widehat R_1(t) - \widehat R_0 \right ) dt
=
 \frac{1}{K}
\sum_{k=1}^K 
\frac{1}{K-1} \left ( \frac{\Gamma}{K} - \frac{1}{C_k}  \right )  \frac{ \bE{\sigma} }{C_k} \cdot \bE{\sigma}
\label{eq:IntegratingDifference}
\end{equation}
upon noting that
\begin{equation}
\int_{-\infty}^0 \bP{ C_{k}t + \sigma > 0 } dt
= \frac{1}{C_k} \int_{0}^\infty \bP{ \sigma > x } dx 
= \frac{ \bE{\sigma} }{C_k},
\quad k=1, \ldots , K
\label{eq:AnIdentityExpectation}
\end{equation}
by a simple change of variable.
Uninteresting algebra on (\ref{eq:IntegratingDifference}) readily yield 
(\ref{eq:ExpectedR_k=1}) with the help of 
(\ref{eq:ExpectedR_0Alternate}), and this completes
the proof of Proposition \ref{prop:ExpectedR_k=1} .
\myendpf

\section{The case $n=2$}
\label{sec:n=2}

The computation of the second derivative is given under the following scenario:
The tagged job arrives at time $t=0$ with characteristic pair $(\sigma_0, \Sigma_0)$.
With $s$ and $t$ in $\mathbb{R}$, in addition
to the tagged job, exactly two jobs
arrive over the entire horizon $(-\infty, \infty)$,
say at times $s$ and $t$ with characteristic pairs
$(\sigma_s, \Sigma_s)$ and $(\sigma_t, \Sigma_t)$, respectively.
The tagged job, the $s$-job and the $t$-job are assigned to their
respective servers $\nu_0$ (in $\Sigma_0$), $\nu_s$ (in $\Sigma_s$)
and $\nu_t$ (in $\Sigma_t$)
in accordance with the power-of-two load balancing scheduling policy.

\subsection{Evaluating $\widehat R_2(s,t)$}

For each $s$ and $t$ in $\mathbb{R}$, we have
\begin{equation}
\widehat R_2(s,t) = \bE{ R_2(s,t) }
\quad
\mbox{with~}
R_2(s,t) = \frac{ \sigma_0 }{ C_{\nu_0} } .
\label{eq:n=2}
\end{equation}
The server assignment rvs
$\nu_0$, $\nu_s$ and $\nu_t$ do not always coincide
with $\nu^\star_0$, $\nu^\star_s$ and $\nu^\star_t$, respectively,
because these rvs 
may be affected by whether earlier jobs
have completed service by the time server selection needs to be determined.

\begin{proposition}
{\sl Under the enforced independence assumptions,
we have $\widehat R_2(s,t) = \widehat R_0$ for $0 < s < t$ and $\widehat R_2(s,t) = \widehat R_1(s)$ for $s < 0 < t$,
while for $ s < t < 0 $, it holds that
\begin{eqnarray}
\widehat R_2(s,t)
&=&
\left (
 \frac{1}{K} \sum_{k=1}^K \bP{ s + \frac{\sigma_s}{C_k} \leq t }
\right )
\cdot 
\widehat R_1 (t)
\nonumber \\
& &
~ +
\frac{1}{K(K-1)}
\sum_{k=1}^K \left (
\sum_{\ell=1, \ell \neq k}^K \bP{ t < s + \frac{\sigma_s}{C_\ell} \leq 0 }
\right )
\cdot H_k(t)
\nonumber \\
& &
~+
\frac{1}{K(K-1)^2}
\sum_{k=1}^K \sum_{\ell=1, \ell \neq k}^K
\left ( \Gamma - \frac{1}{C_\ell} \right )
\bP{C_\ell s + \sigma_s > 0 } \bP{ C_k t + \sigma_t \leq  0 }
\cdot \bE{\sigma_0}
\nonumber \\
& &
~+
\frac{1}{K^2(K-1)^2}
\sum_{k=1}^K
\sum_{\ell=1, \ell \neq k }^K  \Sigma_{k\ell}
\bP{  C_k s+ \sigma_s > 0 } \bP{ C_\ell t + \sigma_t > 0 }
\cdot \bE{\sigma_0}
\label{eq:ExpectedR_2C}
\end{eqnarray}
with
\begin{equation}
\Sigma_{k\ell}
=
(K+1) \Gamma - K \left ( \frac{1}{C_k} + \frac{1}{C_\ell} \right ),
\quad k,\ell =1, \ldots , K.
\label{eq:Sigma_kl}
\end{equation}
}
\label{prop:ExpectedR_2}
\end{proposition}

Before starting the proof of Proposition \ref{prop:ExpectedR_2}
in Section \ref{sec:ProofPropExpectedR_2}, we pause to give a more compact expression for (\ref{eq:ExpectedR_2C}).

\subsection{Towards a more compact expression for (\ref{eq:ExpectedR_2C}) }

As we focus on the  last two terms in (\ref{eq:ExpectedR_2C}),
interchange the dummy indices $k$ and $\ell$, and then change the order of summations in the resulting expression. 
We can readily check that
\begin{eqnarray}
& & \frac{1}{K(K-1)^2}
\sum_{k=1}^K \sum_{\ell=1, \ell \neq k}^K
\left ( \Gamma - \frac{1}{C_\ell} \right )
\bP{C_\ell s + \sigma_s > 0 } \bP{ C_k t + \sigma_t \leq  0 }
\cdot \bE{\sigma_0}
\nonumber \\
& &
~+
\frac{1}{K^2(K-1)^2}
\sum_{k=1}^K
\sum_{\ell=1, \ell \neq k }^K  \Sigma_{k\ell}
\cdot
\bP{  C_k s+ \sigma_s > 0 } \bP{ C_\ell t + \sigma_t > 0 }
\cdot \bE{\sigma_0}
\nonumber \\
&=& 
\frac{1}{K^2(K-1)^2}
\sum_{k=1}^K
\sum_{\ell=1, \ell \neq k }^K  G_{k\ell}(s,t) \cdot \bE{\sigma_0}
\end{eqnarray}
with
\begin{eqnarray}
G_{k\ell}(s,t) 
&=&
K \left ( \Gamma - \frac{1}{C_k} \right )
\bP{C_k s + \sigma_s > 0 } \bP{ C_\ell t + \sigma_t \leq  0 }
\nonumber \\
& &
~+
\left ( (K+1) \Gamma - K \left ( \frac{1}{C_k} + \frac{1}{C_\ell} \right ) \right )
\cdot
\bP{  C_k s+ \sigma_s > 0 } \bP{ C_\ell t + \sigma_t > 0 }
\nonumber \\
&=&
K \left ( \Gamma - \frac{1}{C_k} \right )
\bP{C_k s + \sigma_s > 0 } 
\nonumber \\
& &
~+
K \left (  \frac{\Gamma}{K} -   \frac{1}{C_\ell} \right )
\bP{  C_k s+ \sigma_s > 0 } \bP{ C_\ell  t + \sigma_t > 0 }
\end{eqnarray}
for every $k,\ell =1, \ldots ,K$.
Upon substitution into (\ref{eq:ExpectedR_2C}), we then conclude that
\begin{eqnarray}
\widehat R_2(s,t)
&=&
\left (
 \frac{1}{K} \sum_{k=1}^K \bP{ s + \frac{\sigma_s}{C_k} \leq t }
\right )
\cdot 
\widehat R_1 (t)
\nonumber \\
& &
~ +
\frac{1}{K(K-1)}
\sum_{k=1}^K \left (
\sum_{\ell=1, \ell \neq k}^K \bP{ t < s + \frac{\sigma_s}{C_\ell} \leq 0 }
\right )
\cdot H_k(t)
\nonumber \\
& &
~+ 
\frac{1}{K(K-1)^2}
\sum_{k=1}^K
\sum_{\ell=1, \ell \neq k }^K 
\left (  \frac{\Gamma}{K} -   \frac{1}{C_\ell} \right )
\bP{  C_k s+ \sigma_s > 0 } \bP{ C_\ell  t + \sigma_t > 0 }
\cdot \bE{\sigma_0 }
\nonumber \\
& &
~+ 
\frac{1}{K(K-1)}
\sum_{k=1}^K
\left ( \Gamma -   \frac{1}{C_k} \right )
\bP{  C_k s+ \sigma_s > 0 } 
\cdot \bE{\sigma_0}.
\label{eq:ExpectedR_2C+X}
\end{eqnarray}

Next using (\ref{eq:H_k(t)_Alternate}) we get
\begin{eqnarray}
& &
\sum_{k=1}^K \left (
\sum_{\ell=1, \ell \neq k}^K \bP{ t < s + \frac{\sigma_s}{C_\ell} \leq 0 }
\right )
\cdot H_k(t)
\nonumber \\
&=& 
\sum_{k=1}^K \left (
\sum_{\ell=1, \ell \neq k}^K \bP{ t < s + \frac{\sigma_s}{C_\ell} \leq 0 }
\right )
\left (
\widehat R_0
+
\frac{1}{K-1} \left ( \frac{\Gamma}{K} - \frac{1}{C_k}  \right )
\bP{ C_k t + \sigma_t > 0 } \cdot \bE{\sigma_0}
\right )
\nonumber
\end{eqnarray}
and the second term in (\ref{eq:ExpectedR_2C+X}) becomes
\begin{eqnarray}
& & 
\frac{1}{K(K-1)}
\sum_{k=1}^K \left (
\sum_{\ell=1, \ell \neq k}^K \bP{ t < s + \frac{\sigma_s}{C_\ell} \leq 0 }
\right )
\cdot H_k(t)
\nonumber \\
&=& 
\left (
\frac{1}{K}
\sum_{\ell=1}^K \bP{ t < s + \frac{\sigma_s}{C_\ell} \leq 0 }
\right ) \widehat R_0
\nonumber \\
& &
~+ 
\frac{1}{K(K-1)^2}
\sum_{k=1}^K 
\sum_{\ell=1, \ell \neq k }^K 
\left ( \frac{\Gamma}{K} - \frac{1}{C_k}  \right )
\bP{ t < s + \frac{\sigma_s}{C_\ell} \leq 0 }
\bP{ C_k t + \sigma_t > 0 } \cdot \bE{\sigma_0}.
\nonumber
\end{eqnarray}

Substituting this last expression into (\ref{eq:ExpectedR_2C+X}) we readily get
the following more compact expression for (\ref{eq:ExpectedR_2C}).

\begin{proposition}
{\sl Under the enforced independence assumptions,
for $ s < t < 0 $, it holds that
\begin{eqnarray}
\widehat R_2(s,t)
&=&
\left (
 \frac{1}{K} \sum_{\ell=1}^K \bP{ s + \frac{\sigma_s}{C_\ell} \leq t }
\right )
\cdot 
\widehat R_1 (t)
+
\left ( \frac{1}{K}
\sum_{\ell=1}^K \bP{ t < s + \frac{\sigma_s}{C_\ell} \leq 0 }
\right )
\cdot \widehat R_0
\nonumber \\
& &
~ 
+ \frac{1}{K(K-1)}
\sum_{k=1}^K
\left ( \Gamma -   \frac{1}{C_k} \right )
\bP{  C_k s+ \sigma_s > 0 } 
\cdot \bE{\sigma_0}  +  \frac{1}{K(K-1)^2}  \cdot H(s,t)
\nonumber 
\end{eqnarray}
where we have set
\begin{eqnarray}
H(s,t)
=
\sum_{\ell=1}^K
\sum_{k=1, k \neq \ell}^K 
\left ( \frac{\Gamma}{K} - \frac{1}{C_\ell}  \right )
\bP{ C_k t < C_k s + \sigma_s }
\bP{ C_\ell t + \sigma_t > 0 } \cdot \bE{\sigma_0}.
\nonumber
\end{eqnarray}
}
\label{prop:ExpectedR_2MoreCompact}
\end{proposition}

\section{A proof of Proposition \ref{prop:ExpectedR_k=2}}.

Our point of departure is the expression (\ref{eq:ExpressionFor_n=2}).
For notational simplicity we shall write
\[
R^\star (s,t)
= 
\widehat R_2(s,t) -  \widehat R_1(s) - \widehat R_1(t) +  \widehat R_0,
\quad s,t \in \mathbb{R}.
\]

\subsection{The integral to be evaluated}

We start with
\begin{eqnarray}
R^{\prime\prime}(0+)
&=&
\int_{\mathbb{R}}
\left (  \int_{\mathbb{R}} R^\star (s,t) dt \right )
ds
\nonumber \\
&=&
\int_{\mathbb{R}}
\left (  \int_{-\infty}^s R^\star (s,t) dt \right )
ds
+
\int_{\mathbb{R}}
\left ( \int_{s}^\infty R^\star (s,t) dt \right )
ds.
\label{eq:DecompositionIntegral}
\end{eqnarray}

The second term in this expression can be written as
\begin{equation}
\int_{\mathbb{R}}
\left (  \int_{s}^\infty R^\star (s,t) dt \right )
ds
=
\int_{-\infty}^0
\left (  \int_{s}^\infty R^\star (s,t) dt \right )
ds
+
\int_{0}^\infty
\left (  \int_{s}^\infty R^\star (s,t) dt \right )
ds .
\label{eq:DecompositionA}
\end{equation}
Now, by Propositions \ref{prop:ExpectedR_1} and \ref{prop:ExpectedR_2} we have
$ R^\star (s,t)  = \widehat R_0 - \widehat R_0  - \widehat R_0  + \widehat R_0 = 0$ 
whenever $ 0 < s < t $, and the conclusion
\begin{equation}
\int_{0}^\infty
\left (  \int_{s}^\infty R^\star (s,t) dt \right )
ds = 0
\label{eq:DecompositionB}
\end{equation}
follows.

Next, consider the decomposition 
\begin{equation}
\int_{-\infty}^0
\left (  \int_{s}^\infty R^\star (s,t) dt \right )
ds
=
\int_{-\infty}^0
\left (  \int_{s}^0 R^\star (s,t) dt \right )
ds
+
\int_{-\infty}^0
\left (  \int_{0}^\infty R^\star (s,t) dt \right )
ds.
\label{eq:DecompositionC}
\end{equation}
On the range $s < 0 < t$, 
Propositions \ref{prop:ExpectedR_1} and \ref{prop:ExpectedR_2} yield
$\widehat R_2(s,t) = \widehat R_1(s)$ and 
$\widehat R_1(t) = \widehat R_0 $, whence
$R^\star (s,t)  = \widehat R_1(s)  - \widehat R_1(s)  - \widehat R_0  + \widehat R_0  = 0$ again, so that
\[
\int_{-\infty}^0
\left (  \int_{0}^\infty R^\star (s,t) dt \right ) ds = 0.
\]
Combining (\ref{eq:DecompositionA}), (\ref{eq:DecompositionB}) and (\ref{eq:DecompositionC}),
we conclude that the second term in (\ref{eq:DecompositionIntegral}) reduces to
\begin{equation}
\int_{\mathbb{R}}
\left (  \int_{s}^\infty R^\star (s,t) dt \right )
ds
=
\int_{-\infty}^0
\left (  \int_{s}^0 R^\star (s,t) dt \right )
ds.
\label{eq:n=2IntegralToBeComputed}
\end{equation}

Finally, returning to the first term in the decomposition (\ref{eq:DecompositionIntegral}) we get
\begin{eqnarray}
\int_{\mathbb{R}}
\left (  \int_{-\infty}^s R^\star (s,t) dt \right )
ds
&=&
\int_{\mathbb{R}}
\left (  \int_{t}^\infty R^\star (s,t) ds \right )
dt
\nonumber \\
&=&
\int_{\mathbb{R}}
\left (  \int_{s}^\infty R^\star (t,s) dt \right )
ds
\nonumber \\
&=&
\int_{\mathbb{R}}
\left (  \int_{s}^\infty R^\star (s,t) dt \right )
ds
\label{eq:n=2IntegralToBeComputedFinal}
\end{eqnarray}
as we note that $R^\star (t,s) = R^\star (s,t)$ for arbitrary $s,t$ in $\mathbb{R}$
since the symmetry $\widehat R_2(t,s)  = \widehat R_2(s,t) $ holds under the enforced statistical assumptions.
It follows from (\ref{eq:DecompositionIntegral}) that
\begin{equation}
R^{\prime\prime}(0+)
=
\int_{\mathbb{R}}
\left (  \int_{\mathbb{R}} R^\star (s,t) dt \right ) ds
= 2 \int_{\mathbb{R}}
\left (  \int_{s}^\infty R^\star (s,t) dt \right ) ds
=
2
\int_{-\infty}^0
\left (  \int_{s}^0 R^\star (s,t) dt \right )
ds
\label{eq:n=2IntegralToBeComputed3}
\end{equation}
as we combine (\ref{eq:DecompositionIntegral}), (\ref{eq:n=2IntegralToBeComputed}) and (\ref{eq:n=2IntegralToBeComputedFinal}).

\subsection{Computing the integrand in (\ref{eq:n=2IntegralToBeComputed3})}

On the way to evaluating the integral (\ref{eq:n=2IntegralToBeComputedFinal}) we consider
$R^\star (s,t) $ for $s < t < 0$. On that range,  
applying (\ref{eq:Integrand}) with $t$ replaced by $s$ yields
\begin{equation}
\widehat R_1(s) - \widehat R_0
=
 \frac{1}{K(K-1)
 }
\sum_{k=1}^K 
\left ( \frac{\Gamma}{K} - \frac{1}{C_k}  \right )
\bP{ C_k s + \sigma_s > 0 } \cdot \bE{\sigma_0}
\label{eq:IntegrandAgain}
\end{equation}
Using the expression for $\widehat R_2(s,t)$ in Proposition \ref{prop:ExpectedR_2MoreCompact} 
and recalling the expression for $\widehat R_0$  we then readily get
\begin{eqnarray}
R^\star (s,t)
&=&
\left ( \widehat R_0 - \widehat R_1(s) \right )
+ \left ( \widehat R_2(s,t) -  \widehat R_1(t)  \right )
\nonumber \\
&=&
-  \frac{1}{K(K-1)}
\sum_{k=1}^K 
\left ( \frac{\Gamma}{K} - \frac{1}{C_k}  \right )
\bP{ C_k s + \sigma_s > 0 } \cdot \bE{\sigma_0}
\nonumber \\
& &
+
\left ( \frac{1}{K}
\sum_{\ell=1}^K \bP{ t < s + \frac{\sigma_s}{C_\ell} \leq 0 }
\right )
\cdot \widehat R_0 
+
\left ( 
\frac{1}{K} \sum_{\ell=1}^K \bP{ s + \frac{\sigma_s}{C_\ell} \leq  t } - 1 
\right )
\cdot 
\widehat R_1 (t) 
\nonumber \\
& &
~ + \frac{1}{K(K-1)}
\sum_{k=1}^K
\left (  \Gamma -   \frac{1}{C_k} \right )
\bP{  C_k s+ \sigma_s > 0 } 
\cdot \bE{\sigma_0} +  \frac{1}{K(K-1)^2}  \cdot H(s,t)
\nonumber \\
&=&
\left ( \frac{1}{K}
\sum_{\ell=1}^K \bP{ t < s + \frac{\sigma_s}{C_\ell} \leq 0 }
\right )
\cdot \widehat R_0 
-  \left (
\frac{1}{K} \sum_{\ell=1}^K \bP{ s + \frac{\sigma_s}{C_\ell} >  t }
\right )
\cdot
\widehat R_1 (t)
\nonumber \\
& &
~ +
\frac{1}{K}
\sum_{k=1}^K 
\bP{ C_k s + \sigma_s > 0 } \cdot \left ( \frac{\Gamma}{K} \bE{\sigma_0} \right )
+  \frac{1}{K(K-1)^2}  \cdot H(s,t)
\nonumber \\
&=&
\left ( \frac{1}{K}
\sum_{k=1}^K \bP{ C_k t < C_k s + \sigma_s }
\right )
\cdot \widehat R_0
-  \left (
\frac{1}{K} \sum_{k=1}^K \bP{ s + \frac{\sigma_s}{C_k} >  t }
\right )
\cdot \widehat R_1 (t)
\nonumber \\
& & 
+  \frac{1}{K(K-1)^2}  \cdot H(s,t)
\nonumber \\
&=&
\left ( \frac{1}{K}
\sum_{k=1}^K \bP{ C_k t < C_k s + \sigma_s }
\right )
\cdot \left ( \widehat R_0 -\widehat R_1 (t) \right )
+  \frac{1}{K(K-1)^2}  \cdot H(s,t) .
\label{eq:For_n=2IntegrandToBeComputed}
\end{eqnarray}

\subsection{Evaluating (\ref{eq:n=2IntegralToBeComputed3})}

Next, recall that in these expressions the rvs
$\sigma_s$ and $\sigma_0$ are distributed like $\sigma$.
Thus, after a change of order of integration and a change of variable, we note that
\begin{eqnarray}
& & \int_{-\infty}^0
\left (  \int_{s}^0 
\left ( \frac{1}{K}
\sum_{k=1}^K \bP{ C_k t < C_k s + \sigma }
\right )
\cdot \left ( \widehat R_0 -\widehat R_1 (t) \right )
dt \right ) ds
\nonumber \\
&=&
\frac{1}{K}
\sum_{k=1}^{K} 
\int_{-\infty}^0
\left (  \int_{s}^0
\bP{ C_k t < C_k s + \sigma }
\cdot \left ( \widehat R_0 -\widehat R_1 (t) \right )
dt \right ) ds
\nonumber \\
&=&
\frac{1}{K}
\sum_{k=1}^{K} 
\int_{-\infty}^0
\left (  \int_{-\infty}^t
\bP{ C_k t < C_k s + \sigma }
\cdot \left ( \widehat R_0 -\widehat R_1 (t) \right )
ds \right ) dt
\nonumber \\
&=&
\frac{1}{K}
\sum_{k=1}^{K} 
\int_{-\infty}^0
\left (  \int_{-\infty}^t \bP{ C_k t < C_k s + \sigma } ds \right )
\cdot \left ( \widehat R_0 -\widehat R_1 (t) \right )
dt
\nonumber \\
&=&
\frac{1}{K}
\sum_{k=1}^{K} 
\int_{-\infty}^0
\left (  \int_0^{\infty} \bP{ C_k x < \sigma } dx \right )
\cdot \left ( \widehat R_0 -\widehat R_1 (t) \right )
dt
\nonumber \\
&=&
\frac{1}{K}
\sum_{k=1}^{K} 
\int_{-\infty}^0
\frac{ \bE{\sigma} } {C_k}
\cdot \left ( \widehat R_0 -\widehat R_1 (t) \right )
dt
\nonumber \\
&=&
- \left ( 
\int_{-\infty}^0
\left ( \widehat R_1 (t) -  \widehat R_0 \right )
dt 
\right ) 
\cdot \frac{\Gamma}{K} \bE{\sigma}
\nonumber \\
&=&
-
\left ( \frac{1}{K(K-1)}
\sum_{k=1}^K 
 \left ( \frac{\Gamma}{K} - \frac{1}{C_k}  \right )  \frac{ \bE{\sigma} }{C_k} \cdot \bE{\sigma}
\right ) 
\cdot \frac{\Gamma}{K} \bE{\sigma}
\nonumber \\
&=&
\left ( 
\frac{1}{K} \sum_{k=1}^K \frac{ 1 }{C^2_k} - \left ( \frac{ \Gamma}{K} \right )^2 
\right )
\cdot  \frac{\Gamma}{K(K-1)} \left ( \bE{\sigma } \right )^3
\label{eq:For_n=2+A}
\end{eqnarray}
where the step before last made used of the expression (\ref{eq:IntegratingDifference}).

In a similar vein, we find that
\begin{eqnarray}
\int_{-\infty}^0
\left (  \int_{s}^0 H (s,t) dt \right ) ds
=
\frac{1}{K(K-1)^2} 
\sum_{\ell=1}^K
\sum_{k=1, k \neq \ell}^K 
\left ( \frac{\Gamma}{K} - \frac{1}{C_\ell}  \right )  I_{k\ell}  \cdot \bE{\sigma}
\label{eq:For_n=2+B}
\end{eqnarray}
with
\begin{eqnarray}
I_{k\ell}
&=&
\int_{-\infty}^0
\left (  \int_{s}^0
\bP{ C_k t < C_k s + \sigma }
\bP{ C_\ell t + \sigma > 0 }
dt \right ) ds
\nonumber \\
&=&
\int_{-\infty}^0
\left (  \int_{-\infty}^t
\bP{ C_k t < C_k s + \sigma  }
ds \right )  \bP{ C_\ell t + \sigma > 0 } dt
\nonumber \\
&=&
\int_{-\infty}^0
\left (  \int_{0}^\infty \bP{ C_k x <  \sigma } dx \right )  \bP{ C_\ell t + \sigma > 0 } dt
\nonumber \\
&=&
\left (  \int_{0}^\infty \bP{ C_k x <  \sigma } dx \right )  
\left ( \int_{-\infty}^0 \bP{ C_\ell t + \sigma > 0 } dt \right )
\nonumber \\
&=&
\frac{ \bE{\sigma} }{C_k} \cdot \frac{ \bE{\sigma} }{C_\ell},
\quad k,\ell =1, \ldots , K.
\label{eq:For_n=2+Ba}
\end{eqnarray}

Therefore,
\begin{equation}
\int_{-\infty}^0
\left (  \int_{s}^0 H (s,t) dt \right ) ds
=
\sum_{\ell=1}^K
\sum_{k=1, k \neq \ell}^K 
\left ( \frac{\Gamma}{K} - \frac{1}{C_\ell}  \right ) 
\left (
\frac{ \bE{\sigma} }{C_k} \cdot \frac{ \bE{\sigma} }{C_\ell}
\right )
\cdot \bE{\sigma}
\label{eq:For_n=2+Bb}
\end{equation}
with
\begin{eqnarray}
\sum_{\ell=1}^K
\sum_{k=1, k \neq \ell}^K \frac{ \bE{\sigma} }{C_k} \cdot \frac{ \bE{\sigma} }{C_\ell}
&=&
\sum_{\ell=1}^K  \frac{ \bE{\sigma} }{C_\ell} 
\left ( \sum_{k=1, k \neq \ell}^K \frac{ \bE{\sigma} }{C_k} \right )
\nonumber \\
&=&
\sum_{\ell=1}^K  \frac{ 1 }{C_\ell} 
\left ( \Gamma - \frac{1}{ C_\ell}  \right ) \cdot \left ( \bE{\sigma} \right )^2
\nonumber \\
&=&
\left (
\Gamma^2  -  \sum_{\ell=1}^K  \frac{ 1 }{C^2_\ell} 
\right ) 
\cdot \left ( \bE{\sigma} \right )^2
\label{eq:For_n=2+Bc}
\end{eqnarray}
and
\begin{eqnarray}
\sum_{\ell=1}^K
\sum_{k=1, k \neq \ell}^K 
\frac{1}{C_\ell} 
\left ( \frac{ \bE{\sigma} }{C_k} \cdot \frac{ \bE{\sigma} }{C_\ell} \right )
&=&
\sum_{\ell=1}^K  \frac{ \bE{\sigma} }{C^2_\ell} 
\left ( \sum_{k=1, k \neq \ell}^K \frac{ \bE{\sigma} }{C_k} \right )
\nonumber \\
&=&
\sum_{\ell=1}^K  \frac{ 1 }{C^2_\ell} 
\left ( \Gamma - \frac{1}{ C_\ell}  \right ) \cdot \left ( \bE{\sigma} \right )^2
\nonumber \\
&=&
\left (
\Gamma \sum_{\ell=1}^K \frac{1}{C^2_\ell}   -  \sum_{\ell=1}^K  \frac{ 1 }{C^3_\ell} 
\right ) 
\cdot \left ( \bE{\sigma} \right )^2.
\label{eq:For_n=2+Bd}
\end{eqnarray}
Substitute (\ref{eq:For_n=2+Bc}) and (\ref{eq:For_n=2+Bd}) into (\ref{eq:For_n=2+Bb}), and we find
\begin{eqnarray}
\lefteqn{ \int_{-\infty}^0 \left (  \int_{s}^0 H (s,t) dt \right ) ds }
& & 
\nonumber \\
&=&
\frac{1}{K(K-1)^2} 
\left (
\frac{\Gamma}{K}
\left (
\Gamma^2  -  \sum_{\ell=1}^K  \frac{ 1 }{C^2_\ell} 
\right ) 
-
\left (
\Gamma \sum_{\ell=1}^K \frac{1}{C^2_\ell}   -  \sum_{\ell=1}^K  \frac{ 1 }{C^3_\ell} 
\right ) 
\right )
\cdot \left ( \bE{\sigma} \right )^3
\nonumber \\
&=&
\frac{1}{K(K-1)^2} 
\left (
\frac{\Gamma^3}{K}
- \frac{K+1}{K} \cdot \Gamma \sum_{\ell=1}^K  \frac{ 1 }{C^2_\ell} 
+ \sum_{\ell=1}^K  \frac{ 1 }{C^3_\ell} 
\right )
\cdot \left ( \bE{\sigma} \right )^3.
\label{eq:For_n=2+C}
\end{eqnarray}

Finally, return to (\ref{eq:For_n=2IntegrandToBeComputed}) and collect (\ref{eq:For_n=2+A}) and (\ref{eq:For_n=2+C}): Uninteresting calculations
show that
\begin{eqnarray}
\lefteqn{ \int_{-\infty}^0 \left (  \int_{s}^0 R^\star_2 (s,t) dt \right ) ds }
& & 
\nonumber \\
&=&
\left ( 
- \left ( \frac{ \Gamma}{K} \right )^2 
+ \frac{1}{K} \sum_{k=1}^K \frac{ 1 }{C^2_k}
\right )
\cdot  \frac{\Gamma}{K(K-1)} \left ( \bE{\sigma } \right )^3
\nonumber \\
& &
~ +
\frac{1}{K(K-1)^2} 
\left (
\frac{\Gamma^3}{K}
- \frac{K+1}{K} \Gamma \sum_{\ell=1}^K  \frac{ 1 }{C^2_\ell} 
+ \sum_{\ell=1}^K  \frac{ 1 }{C^3_\ell} 
\right )
\left ( \bE{\sigma } \right )^3
\nonumber \\
&=&
\left ( -  \frac{1}{K^3(K-1)}  + \frac{1}{K^2(K-1)^2}  \right ) \Gamma^3 \cdot \left ( \bE{\sigma } \right )^3
\nonumber \\
& &
~ +
\left ( \frac{1}{K^2 (K-1)}  - \frac{ K+1 }{ K^2(K-1)^2} \right )
\left ( \sum_{k=1}^K \frac{ 1 }{C^2_k} \right ) 
\Gamma \cdot \left ( \bE{\sigma } \right )^3
 \nonumber \\
& &
~ +
\frac{1}{K(K-1)^2}
\left ( \sum_{\ell=1}^K  \frac{ 1 }{C^3_\ell} 
\right )
\cdot \left ( \bE{\sigma } \right )^3
\nonumber \\
&=&
\frac{1}{(K-1)^2}
\left (
 \left ( \frac{\Gamma}{K} \right )^3
- 2
\left ( \frac{1}{K} \sum_{k=1}^K \frac{ 1 }{C^2_k} \right ) 
\left ( \frac{ \Gamma }{K} \right ) 
+
\frac{1}{K}  \sum_{\ell=1}^K  \frac{ 1 }{C^3_\ell} 
\right ) 
\cdot \left ( \bE{\sigma } \right )^3,
\end{eqnarray}
and the expression (\ref{eq:ExpectedR_k=2}) now follows from (\ref{eq:n=2IntegralToBeComputed3}).
\myendpf

\section{A proof of Proposition \ref{prop:ExpectedR_2} }
\label{sec:ProofPropExpectedR_2}

The cases $0 < s< t $
and $ s< 0 < t$ are straightforward by virtue of the operational
assumptions of the power-of-two load balancing policy.
Indeed, when $0 < s< t $, $\nu_0 = \nu^\star_0$, hence
$R_2(s,t) = R_0$ and $\widehat R_2(s,t) = \widehat R_0$ holds.
On the other hand, when $s< 0 < t$, the future $t$-job 
does not affect the selection of $\nu_0$, hence has no impact on
the performance of the tagged customer. As only the $s$-job 
can possibily affect the choice of $\nu_0$, we get
$R_2(s,t) = R_1(s)$ and this shows that $\widehat R_2(s,t) = \widehat R_1(s)$.

From now on we assume $s < t <  0$, in which case we have
$\nu_s = \nu^\star_s$.
The selection of $\nu_t$ can
in principle be affected by whether the $s$-job has completed 
its service by time $t$, while that of $\nu_0$ 
will be determined by whether the $s$-job and  $t$-job have
completed service by the time the tagged job enters the system.
Therefore, as the $s$-job 
completes at time $ s + \frac{\sigma_s}{C_{\nu^\star_s}}$,
several possibilities arise; they are  captured in the decomposition
\begin{eqnarray}
\bE{ R_2(s,t) }
&=& 
\bE{ \1{ s + \frac{\sigma_s}{C_{\nu^\star_s}} \leq t} R_2(s,t) }
\nonumber \\
& &
~+ \bE{ \1{ t < s + \frac{\sigma_s}{C_{\nu^\star_s}} \leq 0 }
R_2(s,t) }
\nonumber \\
& &
~+ \bE{ \1{ s + \frac{\sigma_s}{C_{\nu^\star_s}} >  0 }
\1{ t + \frac{\sigma_t}{C_{\nu_t}} \leq  0 } R_2(s,t) }
\nonumber \\
& &
~+ 
\bE{ \1{ s + \frac{\sigma_s}{C_{\nu^\star_s}} > 0 }
\1{ t + \frac{\sigma_t}{C_{\nu_t}} >  0 } R_2(s,t) }.
\label{eq:DECOMPOSITION}
\end{eqnarray}

These four terms are evaluated separately in the next four lemmas.

\begin{lemma}
{\sl
With $s < t < 0$, we have
\begin{equation}
\bE{ \1{ s + \frac{\sigma_s}{C_{\nu^\star_s}} \leq t} R_2(s,t) }
= 
\left (
 \frac{1}{K} \sum_{k=1}^K \bP{ s + \frac{\sigma_s}{C_k} \leq t }
\right )
\cdot 
\widehat R_1 (t)
\label{eq:Case1}
\end{equation}
}
\label{lem:Case1}
\end{lemma}

\myproof
When $s + \frac{\sigma_s}{C_{\nu^\star_s}} \leq t $, 
the $s$-job will have completed service by the time the $t$-job arrives.
Therefore, conditionally on
$s + \frac{\sigma_s}{C_{\nu^\star_s}} \leq t $, it holds that
$R_2(s,t) =_{st} R_1(t)$, whence
\begin{equation}
\bE{ \1{ s + \frac{\sigma_s}{C_{\nu^\star_s}} \leq t} R_2(s,t) }
= 
\bE{ \1{ s + \frac{\sigma_s}{C_{\nu^\star_s}} \leq t} R_1(t) }
= 
\bP{ s + \frac{\sigma_s}{C_{\nu^\star_s}} \leq t }\cdot \widehat R_1 (t)
\end{equation}
with
\[
\bP{ s + \frac{\sigma_s}{C_{\nu^\star_s}} \leq t }
= \frac{1}{K} \sum_{k=1}^K \bP{ s + \frac{\sigma}{C_k} \leq t }
\]
by the usual arguments.
This completes the proof of (\ref{eq:Case1}).
\myendpf

\begin{lemma}
{\sl
With $s < t < 0$, we have
\begin{eqnarray}
\lefteqn{
\bE{ \1{ t < s + \frac{\sigma_s}{C_{\nu^\star_s}} \leq 0 } R_2(s,t) }
} & &
\nonumber \\
&=&
\frac{1}{K(K-1)}
\sum_{k=1}^K \left (
\sum_{\ell=1, \ell \neq k}^K \bP{ t < s + \frac{\sigma_s}{C_\ell} \leq 0 }
\right )
\cdot H_k(t)
\label{eq:Case2A}
\end{eqnarray}
with $H_k(t)$ given by (\ref{eq:H_k(t)}) for all $k=1, \ldots , K$.
}
\label{lem:Case2A}
\end{lemma}

\myproof
When $t < s + \frac{\sigma_s}{C_{\nu^\star_s}}  \leq 0$,
the $s$-job has not completed its service by time $t$, 
but will have completed it by the time the tagged job arrives.
Thus, only the $t$-job can affect the definition of $\nu_0$
(through $\sigma_t$ and $\nu_t$). 

With this in mind, consider the decomposition
\begin{equation}
\bE{ \1{ t < s + \frac{\sigma_s}{C_{\nu^\star_s}} \leq 0 } R_2(s,t) }
=
\sum_{k=1}^K
\bE{ \1{ t < s + \frac{\sigma_s}{C_{\nu^\star_s}} \leq 0}
\1{ \nu_t = k }
\frac{ \sigma_0 }{C_{\nu_0}} } .
\label{eq:BasicRelationForN=2PieceA}
\end{equation}
Fix $k=1, \ldots , K$.
We are in the setting of Lemma \ref{lem:Auxiliary}
with $y=t$, $E = [ t < s + \frac{\sigma_s}{C_{\nu^\star_s}} \leq 0 ]$,
$\tau = \sigma_t$ and $\gamma = \nu_t$ so that $\gamma_0 = \nu_0$:
The expression (\ref{eq:Auxiliary}) becomes
\begin{eqnarray}
\lefteqn{
\bE{ \1{ t < s + \frac{\sigma_s}{C_{\nu^\star_s}} \leq 0}
\1{\nu_t  = k} \frac{\sigma_0}{C_{\nu_0}} } 
} & &
\nonumber \\
&=&
\bP{ t < s + \frac{\sigma_s}{C_{\nu^\star_s}} \leq 0, 
    \nu_t  = k, t + \frac{\sigma_t}{C_k} \leq 0 }
\cdot
\widehat R_0
\nonumber \\
& &
~+ \frac{1}{K-1}
\bP{ t < s + \frac{\sigma_s}{C_{\nu^\star_s}} \leq 0,
    \nu_t = k, t + \frac{\sigma_t}{C_k} > 0 }
\left ( \Gamma - \frac{1}{C_k} \right ) \cdot \bE{\sigma_0} 
\nonumber \\
&=&
\bP{ t < s + \frac{\sigma_s}{C_{\nu^\star_s}} \leq 0,  \nu_t  = k} H_k(t)
\label{eq:AppB+0}
\end{eqnarray}
with $H_k(t)$ defined at (\ref{eq:H_k(t)}).
In the last step we used the fact that under the enforced independence assumptions,
the rv $\sigma_t$ is independent of the rvs $\{ \sigma_s, \nu^\star_s, \nu_t \}$ 
when $\nu_t$ is generated by the power-of-two load balancing policy.

In Appendix C we show that
\begin{eqnarray}
\bP{ t < s + \frac{\sigma_s}{C_{\nu^\star_s}} \leq 0,  \nu_t  = k}
=
\frac{1}{K(K-1)}
\sum_{\ell=1, \ell \neq k}^K \bP{ t < s + \frac{\sigma_s}{C_\ell} \leq 0 }.
\label{eq:AppB+1}
\end{eqnarray}
Inserting (\ref{eq:AppB+1}) back into
(\ref{eq:AppB+0}) yields
\[
\bE{ \1{ t < s + \frac{\sigma_s}{C_{\nu^\star_s}} \leq 0 }
     \1{ \nu_t  = k } \frac{\sigma_0}{c_{\nu_0}} }
=
\left ( \frac{1}{K(K-1)}
\sum_{\ell=1, \ell \neq k}^K \bP{ t < s + \frac{\sigma_s}{C_\ell} \leq 0 }
\right )
\cdot H_k(t),
\]
and the desired result is now obtained by
making use of  (\ref{eq:BasicRelationForN=2PieceA}).
\myendpf

The last two terms in the decomposition (\ref{eq:DECOMPOSITION}) are more cumbersome
to evaluate. Their expressions are given in the next two lemmas whose proofs can
be found in Sections \ref{sec:ProofLemmaCase2B+1}
and  \ref{sec:ProofLemmaCase2B+2}, respectively.

\begin{lemma}
{\sl
With $s < t < 0$, we have
\begin{eqnarray}
\lefteqn{
\bE{ \1{ s + \frac{\sigma_s}{C_{\nu^\star_s}} > 0 }
\1{ t + \frac{\sigma_t}{C_{\nu_t}} \leq  0 } R_2(s,t) }
} & &
\nonumber \\
&=&
\frac{1}{K(K-1)^2}
\sum_{k=1}^K \sum_{\ell=1, \ell \neq k}^K
\left ( \Gamma - \frac{1}{C_\ell} \right )
\bP{C_\ell s + \sigma_s > 0 } \bP{ C_k t + \sigma_t \leq  0 }
\cdot \bE{\sigma_0}
\label{eq:Case2B+1}
\end{eqnarray}
}
\label{lem:Case2B+1}
\end{lemma}

\begin{lemma}
{\sl
With $s < t < 0$, we have
\begin{eqnarray}
\lefteqn{
\bE{ \1{ s + \frac{\sigma_s}{C_{\nu^\star_s}} > 0 }
\1{ t + \frac{\sigma_t}{C_{\nu_t}} >  0 } R_2(s,t) }
} & &
\nonumber \\
&=&
\frac{1}{K^2(K-1)^2}
\sum_{k=1}^K
\sum_{\ell=1, \ell \neq k }^K  \Sigma_{k\ell}
\cdot
\bP{  C_k s+ \sigma_s > 0 } \bP{ C_\ell t + \sigma_t > 0 }
\cdot \bE{\sigma_0}
\label{eq:Case2B+2}
\end{eqnarray}
with the constants $\Sigma_{k\ell}, \ k,\ell =1, \ldots,$ given by (\ref{eq:Sigma_kl}).
}
\label{lem:Case2B+2}
\end{lemma}

\section{A proof of Lemma \ref{lem:Case2B+1}}
\label{sec:ProofLemmaCase2B+1}

We are in the situation when $s < t < 0$.  
If $s + \frac{\sigma_s}{C_{\nu^\star_s}} > 0 $
(hence $s + \frac{\sigma_s}{C_{\nu^\star_s}} > t$), then 
the $s$-job completes its service only after the tagged arrives,
so that both the $s$-job and $t$-job can possibly affect the definition 
of $\nu_0$. If in addition we have $t + \frac{\sigma_t}{C_{\nu_t}} \leq  0$, then
only the $s$-job can affect  the selection $\nu_0$.

In the usual manner we have the decomposition
\begin{eqnarray}
\lefteqn{
\bE{ \1{ s + \frac{\sigma_s}{C_{\nu^\star_s}} > 0 }
\1{ t + \frac{\sigma_t}{C_{\nu_t}} \leq  0 } R_2(s,t) }
} & &
\nonumber \\
&=&
\sum_{k=1}^K
\bE{ \1{ s + \frac{\sigma_s}{C_{\nu^\star_s}} > 0 } 
\1{ t + \frac{\sigma_t}{C_k} \leq  0 }
\1{\nu_t = k} \frac{\sigma_0}{C_{\nu_0}} } 
\nonumber \\
&=&
\sum_{k=1}^K
\sum_{\ell=1, \ell \neq k}^K
\bE{ \1{ \nu^\star_s = \ell } \1{ s + \frac{\sigma_s}{C_\ell } > 0 } 
\1{ t + \frac{\sigma_t}{C_k} \leq  0 }
\1{\nu_t = k} \frac{\sigma_0}{C_{\nu_0}} } .
\label{eq:PieceZD}
\end{eqnarray}

Pick distinct $k,\ell =1, \ldots , K$.
This time we apply Lemma \ref{lem:Auxiliary}
with $y=s$, $E =[ s + \frac{\sigma_s}{C_\ell} > 0,  \nu_t = k, t + \frac{\sigma_t}{C_k} \leq  0 ]$,
$\tau = \sigma_s $ and $\gamma = \nu^\star_s$ (so that $\gamma_0 = \nu_0$).
This leads to
\begin{eqnarray}
\lefteqn{
\bE{ 
\left (
\1{ s + \frac{\sigma_s}{C_\ell } > 0 } 
\1{ t + \frac{\sigma_t}{C_k} \leq  0 }
\1{\nu_t = k} \right )
\1{ \nu^\star_s = \ell }  \frac{\sigma_0}{C_{\nu_0}} } 
} & &
\nonumber \\
&=&
\bP{ s + \frac{\sigma_s}{C_\ell } > 0 ,  t + \frac{\sigma_t}{C_k} \leq  0 , \nu_t = k, \nu^\star_s = \ell ,
s + \frac{\sigma_s}{C_\ell} \leq 0 } 
\cdot
\widehat R_0
\nonumber \\
& &
~+ \frac{1}{K-1}
\bP{ s + \frac{\sigma_s}{C_\ell } > 0 ,  t + \frac{\sigma_t}{C_k} \leq  0 , \nu_t = k, \nu^\star_s = \ell ,
s + \frac{\sigma_s}{C_\ell} > 0 } 
\left ( \Gamma - \frac{1}{C_\ell} \right ) \cdot \bE{\sigma_0} 
\nonumber \\
&=&
\frac{1}{K-1}
\bP{ s + \frac{\sigma_s}{C_\ell } > 0 ,  t + \frac{\sigma_t}{C_k} \leq  0 , \nu_t = k, \nu^\star_s = \ell } 
\left ( \Gamma - \frac{1}{C_\ell} \right ) \cdot \bE{\sigma_0} 
\nonumber \\
&=&
\frac{1}{K-1}
\bP{ t + \frac{\sigma_t}{C_k} \leq  0 }
\bP{ \nu^\star_s = \ell , s + \frac{\sigma_s}{C_\ell } > 0 , \nu_t = k} 
\left ( \Gamma - \frac{1}{C_\ell} \right ) \cdot \bE{\sigma_0}  
\label{eq:PieceZC}
\end{eqnarray}
since the rvs $\sigma_t$ is independent of the collection $\{ \nu^\star_s, \sigma_s, \nu_t \}$ under
the enforced independence assumptions.
Next, we write
\begin{eqnarray}
\lefteqn{
\bP{ \nu^\star_s = \ell, s + \frac{\sigma_s}{C_\ell } > 0 ,   \nu_t = k } 
}
& &
\\
&=&
\bP{ \nu^\star_s = \ell, s + \frac{\sigma_s}{C_\ell } > 0 ,  \ell \in \Sigma_t,  \nu_t = k } 
+ \bP{ \nu^\star_s = \ell, s + \frac{\sigma_s}{C_\ell } > 0 ,   \ell \notin \Sigma_t, \nu_t = k } 
\nonumber
\end{eqnarray}
Taking terms in turn we first get
\begin{eqnarray}
\bP{ \nu^\star_s = \ell, 
s + \frac{\sigma_s}{C_\ell } > 0 ,  \ell \in \Sigma_t,  \nu_t = k } 
&=&
\bP{ \nu^\star_s = \ell, s + \frac{\sigma_s}{C_\ell } > 0 ,  \Sigma_t = \{ \ell, k \},  \nu_t = k } 
\nonumber \\
&=&
\bP{ \nu^\star_s = \ell, s + \frac{\sigma_s}{C_\ell } > 0 ,  \Sigma_t = \{ \ell, k \} } 
\nonumber \\
&=&
\frac{2}{K^2(K-1)} \bP{ s + \frac{\sigma_s}{C_\ell } > 0 }
\label{eq:PieceZA}
\end{eqnarray}
since under the constraint
$s + \frac{\sigma_s}{C_\ell } > 0$, the fact that
$\nu^\star_s$ is an element of  $\Sigma_t$ forces $\nu_t$ to be the other
element in $\Sigma_t$.
In a similar way, under the constraint $s + \frac{\sigma_s}{C_\ell } > 0$,
$\nu^\star_s$ not being in $\Sigma_t$ implies $\nu_t = \nu^\star_t$, and this
leads to
\begin{eqnarray}
\lefteqn{
\bP{ \nu^\star_s = \ell, 
s + \frac{\sigma_s}{C_\ell } > 0 ,  \ell \notin \Sigma_t,  \nu_t = k } 
}
\nonumber \\
&=&
\bP{ \nu^\star_s = \ell, 
s + \frac{\sigma_s}{C_\ell } > 0 ,  \ell \notin \Sigma_t,  \nu^\star_t = k } 
\nonumber \\
&=&
\sum_{a=1, a \neq k, a  \neq \ell}^K
\bP{ \nu^\star_s = \ell, s + \frac{\sigma_s}{C_\ell } > 0 ,  \Sigma_t = \{a,k\}, \nu^\star_t = k } 
\nonumber \\
&=&
\sum_{a=1, a \neq k, a  \neq \ell}^K
\bP{  \nu^\star_s = \ell} \bP{s + \frac{\sigma_s}{C_\ell } > 0 }
\frac{1}{2} \cdot \frac{2}{K(K-1)}
\nonumber \\
&=&
\frac{K-2} {K^2(K-1)} \bP{s + \frac{\sigma_s}{C_\ell } > 0 }.
\label{eq:PieceZB}
\end{eqnarray}

Collecting (\ref{eq:PieceZA}) and (\ref{eq:PieceZB}) gives
\begin{equation}
\bP{ \nu^\star_s = \ell, s + \frac{\sigma_s}{C_\ell } > 0 ,   \nu_t = k } 
= 
\frac{1} {K(K-1)} \bP{s + \frac{\sigma_s}{C_\ell } > 0 },
\end{equation}
and with the help of (\ref{eq:PieceZC}) we conclude that
\begin{eqnarray}
\lefteqn{
\bE{ 
\left (
\1{ s + \frac{\sigma_s}{C_\ell } > 0 } 
\1{ t + \frac{\sigma_t}{C_k} \leq  0 }
\1{\nu_t = k} \right )
\1{ \nu^\star_s = \ell }  \frac{\sigma_0}{C_{\nu_0}} } 
} & &
\nonumber \\
&=&
\frac{1}{K(K-1)^2}
\bP{ t + \frac{\sigma_t}{C_k} \leq  0 }
\bP{s + \frac{\sigma_s}{C_\ell } > 0 }
\left ( \Gamma - \frac{1}{C_\ell} \right ) \cdot \bE{\sigma_0} .
\end{eqnarray}
Inserting this last expression
into (\ref{eq:PieceZD}) we obtain (\ref{eq:Case2B+1}) as desired.
\myendpf

\section{A proof of Lemma \ref{lem:Case2B+2}}
\label{sec:ProofLemmaCase2B+2}

We are in the situation when $s < t < 0$.
If $s+ \frac{\sigma_s}{C_{\nu^\star_s}} > 0$ and $t + \frac{\sigma_t}{C_{\nu_t}} > 0$, 
then $\nu_t$ is determined by the $s$-job and we must have $\nu^\star_s \neq \nu_t$.
When the tagged job arrives, both $\nu_s (=\nu^\star_s)$ and $\nu_t$ would have already been selected,
with both $s$-job and $t$-job still in service when $\nu_0$ needs to be selected.
In order to establish (\ref{eq:Case2B+2}), we begin with the observation that
\begin{eqnarray}
\lefteqn{
 \bE{ \1{s+ \frac{\sigma_s}{C_{\nu^\star_s}} > 0} \1{ t + \frac{\sigma_t}{C_{\nu_t}} > 0 }  R_2(s;t) }
 } & & 
\nonumber \\
&=&
\bE{ \1{s+ \frac{\sigma_s}{C_{\nu^\star_s}} > 0} \1{ t + \frac{\sigma_t}{C_{\nu_t}} > 0 } 
\frac{ \sigma_0 }{ C_{\nu_0}} }
\nonumber \\
&=&
\sum_{k=1}^K
\sum_{\ell=1, \ell \neq k }^K
\bE{ \1{ \nu^\star_s = k} \1{s+ \frac{\sigma_s}{C_{k}} > 0} 
\1{\nu_t = \ell } \1{ t + \frac{\sigma_t}{C_{\ell}} > 0 } 
\frac{ \sigma_0 }{ C_{\nu_0}} }.
\label{eq:ZZ0}
\end{eqnarray}

To take advantage of this decomposition, pick distinct $k,\ell=1, \ldots ,K$.
As we keep in mind whether $\nu^\star_s$ and $\nu_t$ are in $\Sigma_0$, we shall have
to consider four possible cases:
First, if both $\nu^\star_s$ and $\nu_t$ are in $\Sigma_0$, then $\nu_0 = \nu^\star_0$ and we have
\begin{eqnarray}
& &
\bE{ \1{ \nu^\star_s = k} \1{s+ \frac{\sigma_s}{C_{k}} > 0} 
\1{\nu_t = \ell } \1{ t + \frac{\sigma_t}{C_{\ell}} > 0 } \1{ k \in \Sigma_0 , \ell \in \Sigma_0 }
\frac{ \sigma_0 }{ C_{\nu_0}} }
\nonumber \\
&=&
\bE{ \1{ \nu^\star_s = k} \1{s+ \frac{\sigma_s}{C_{k}} > 0} 
\1{\nu_t = \ell } \1{ t + \frac{\sigma_t}{C_{\ell}} > 0 } \1{ \Sigma_0 = \{ k, \ell \} }
\frac{ \sigma_0 }{ C_{\nu^\star_0}} }
\nonumber \\
&=&
\bP{ \nu^\star_s = k, s+ \frac{\sigma_s}{C_{k}} > 0, \nu_t = \ell , t + \frac{\sigma_t}{C_{\ell}} > 0 }
\bP{ \Sigma_0 = \{ k, \ell \} }
\frac{1}{2} \cdot 
\left ( \frac{1}{C_k} + \frac{1}{C_\ell} \right )
\cdot \bE{ \sigma_0 }
\nonumber \\
&=&
\frac{1}{K(K-1)} 
\bP{ \nu^\star_s = k, s+ \frac{\sigma_s}{C_{k}} > 0, \nu_t = \ell , t + \frac{\sigma_t}{C_{\ell}} > 0 }
\left ( \frac{1}{C_k} + \frac{1}{C_\ell} \right )
\cdot \bE{ \sigma_0 }.
\label{eq:ZZ1}
\end{eqnarray}

Next, if $\nu^\star_s$ is not in $\Sigma_0$ but $\nu_t$ is in $\Sigma_0$, 
then $\nu_0$ is the other element in $\Sigma_0$,  and we get
\begin{eqnarray}
& &
\bE{ \1{ \nu^\star_s = k} \1{s+ \frac{\sigma_s}{C_{k}} > 0} 
\1{\nu_t = \ell } \1{ t + \frac{\sigma_t}{C_{\ell}} > 0 } \1{ k \notin \Sigma_0 , \ell \in \Sigma_0 }
\frac{ \sigma_0 }{ C_{\nu_0}} }
\nonumber \\
&=&
\sum_{a=1, a \neq k, a \neq \ell}^K 
\bE{ \1{ \nu^\star_s = k} \1{s+ \frac{\sigma_s}{C_{k}} > 0} 
\1{\nu_t = \ell } \1{ t + \frac{\sigma_t}{C_{\ell}} > 0 }  \1{ \Sigma_0 = \{ \ell, a \} }
\frac{ \sigma_0 }{ C_{\nu_0}} }
\nonumber \\
&=&
\sum_{a=1, a \neq k, a \neq \ell}^K 
\bE{ \1{ \nu^\star_s = k} \1{s+ \frac{\sigma_s}{C_{k}} > 0} 
\1{\nu_t = \ell } \1{ t + \frac{\sigma_t}{C_{\ell}} > 0 }  \1{ \Sigma_0 = \{ \ell, a \} }
\frac{ \sigma_0 }{ C_a} }
\nonumber \\
&=&
\sum_{a=1, a \neq k, a \neq \ell}^K 
\bP{ \nu^\star_s = k, s+ \frac{\sigma_s}{C_{k}} > 0, \nu_t = \ell , t + \frac{\sigma_t}{C_{\ell}} > 0 }
\bP{ \Sigma_0 = \{ \ell, a \} } \cdot \frac{\bE{\sigma_0}}{C_a}
\nonumber \\
&=&
\frac{2}{K(K-1)}
\left (  \sum_{a=1, a \neq k, a \neq \ell}^K   \frac{1}{C_a} \right ) 
\bP{ \nu^\star_s = k, s+ \frac{\sigma_s}{C_{k}} > 0, \nu_t = \ell , t + \frac{\sigma_t}{C_{\ell}} > 0 }
\cdot \bE{\sigma_0}
\nonumber \\
&=&
\frac{2}{K(K-1)}
\left ( \Gamma - \frac{1}{C_k}  - \frac{1}{C_\ell}  \right )
\bP{ \nu^\star_s = k, s+ \frac{\sigma_s}{C_{k}} > 0, \nu_t = \ell , t + \frac{\sigma_t}{C_{\ell}} > 0 } 
\cdot
\bE{\sigma_0} .
\label{eq:ZZ2}
\end{eqnarray}

In a similar way, if  $\nu^\star_s$ is in $\Sigma_0$ but $\nu_t$ is not in $\Sigma_0$, 
then $\nu_0$ is necessarily the other element in $\Sigma_0$, and we get
\begin{eqnarray}
& &
\bE{ \1{ \nu^\star_s = k} \1{s+ \frac{\sigma_s}{C_{k}} > 0} 
\1{\nu_t = \ell } \1{ t + \frac{\sigma_t}{C_{\ell}} > 0 } \1{ k \in \Sigma_0 , \ell \notin \Sigma_0 }
\frac{ \sigma_0 }{ C_{\nu_0}} }
\nonumber \\
&=&
\sum_{b=1, b \neq k, b \neq \ell}^K 
\bE{ \1{ \nu^\star_s = k} \1{s+ \frac{\sigma_s}{C_{k}} > 0} 
\1{\nu_t = \ell } \1{ t + \frac{\sigma_t}{C_{\ell}} > 0 }  \1{ \Sigma_0 = \{ k, b \} }
\frac{ \sigma_0 }{ C_{\nu_0}} }
\nonumber \\
&=&
\sum_{b=1, b \neq k, b \neq \ell}^K 
\bE{ \1{ \nu^\star_s = k} \1{s+ \frac{\sigma_s}{C_{k}} > 0} 
\1{\nu_t = \ell } \1{ t + \frac{\sigma_t}{C_{\ell}} > 0 }  \1{ \Sigma_0 = \{ k, b \} }
\frac{ \sigma_0 }{ C_b} }
\nonumber \\
&=&
\sum_{b=1, b \neq k, b \neq \ell}^K 
\bP{ \nu^\star_s = k, s+ \frac{\sigma_s}{C_{k}} > 0, \nu_t = \ell , t + \frac{\sigma_t}{C_{\ell}} > 0 }
\bP{ \Sigma_0 = \{ b, k \} } \cdot \frac{\bE{\sigma_0}}{C_b}
\nonumber \\
&=&
\frac{2}{K(K-1)}
\left (   \sum_{b=1, b \neq k, b \neq \ell}^K   \frac{1}{C_b} \right ) 
\bP{ \nu^\star_s = k, s+ \frac{\sigma_s}{C_{k}} > 0, \nu_t = \ell , t + \frac{\sigma_t}{C_{\ell}} > 0 }
\cdot \bE{\sigma_0}
\nonumber \\
&=&
\frac{2}{K(K-1)}
\left ( \Gamma - \frac{1}{C_k}  - \frac{1}{C_\ell}  \right )
\bP{ \nu^\star_s = k, s+ \frac{\sigma_s}{C_{k}} > 0, \nu_t = \ell , t + \frac{\sigma_t}{C_{\ell}} > 0 }
\cdot \bE{\sigma_0}.
\label{eq:ZZ3}
\end{eqnarray}

Finally,  when neither $\nu^\star_s$ nor $\nu_t$ are in $\Sigma_0$, then $\nu_0 = \nu^\star_0$, whence
\begin{eqnarray}
& &
\bE{ \1{ \nu^\star_s = k} \1{s+ \frac{\sigma_s}{C_{k}} > 0} 
\1{\nu_t = \ell } \1{ t + \frac{\sigma_t}{C_{\ell}} > 0 } \1{ k \notin \Sigma_0 , \ell \notin \Sigma_0 }
\frac{ \sigma_0 }{ C_{\nu_0}} }
\nonumber \\
&=&
\bE{ \1{ \nu^\star_s = k} \1{s+ \frac{\sigma_s}{C_{k}} > 0} 
\1{\nu_t = \ell } \1{ t + \frac{\sigma_t}{C_{\ell}} > 0 } \1{ k \notin \Sigma_0 , \ell \notin \Sigma_0 }
\frac{ \sigma_0 }{ C_{\nu^\star_0}} }
\nonumber \\
&=&
\sum_{a=1, a \neq k, a \neq \ell}^K
\sum_{b=1, b \neq a, b \neq k, b \neq \ell}^{a-1}
\left ( \ldots \right )_{k\ell}
\end{eqnarray}
with
\begin{eqnarray}
\left ( \ldots \right )_{k\ell}
&=&
\bE{ \1{ \nu^\star_s = k} \1{s+ \frac{\sigma_s}{C_{k}} > 0} 
\1{\nu_t = \ell } \1{ t + \frac{\sigma_t}{C_{\ell}} > 0 } \1{ \Sigma_0 = \{ a,b \} }
\frac{ \sigma_0 }{ C_{\nu^\star_0}} }
\nonumber \\
&=&
\bP{ \nu^\star_s = k, s+ \frac{\sigma_s}{C_{k}} > 0, \nu_t = \ell , t + \frac{\sigma_t}{C_{\ell}} > 0 }
\bP{ \Sigma_0 = \{ a, b \} }
\frac{1}{2} \cdot 
\left ( \frac{1}{C_a} + \frac{1}{C_b} \right )
\cdot \bE{ \sigma_0 }
\nonumber \\
&=&
\frac{1}{K(K-1)}
\bP{ \nu^\star_s = k, s+ \frac{\sigma_s}{C_{k}} > 0, \nu_t = \ell , t + \frac{\sigma_t}{C_{\ell}} > 0 }
\left ( \frac{1}{C_a} + \frac{1}{C_b} \right )
\cdot \bE{ \sigma_0 }.
\label{eq:ZZ4}
\end{eqnarray}
It then follows that
\begin{eqnarray}
& &
\bE{ \1{ \nu^\star_s = k} \1{s+ \frac{\sigma_s}{C_{k}} > 0} 
\1{\nu_t = \ell } \1{ t + \frac{\sigma_t}{C_{\ell}} > 0 } \1{ k \notin \Sigma_0 , \ell \notin \Sigma_0 }
\frac{ \sigma_0 }{ C_{\nu_0}} }
\nonumber \\
&=&
\frac{1}{K(K-1)}
\cdot H_{k\ell} \cdot
\bP{ \nu^\star_s = k, s+ \frac{\sigma_s}{C_{k}} > 0, \nu_t = \ell , t + \frac{\sigma_t}{C_{\ell}} > 0 }
\cdot
\bE{ \sigma_0 }
\label{eq:ZZ5}
\end{eqnarray}
where we have set
\begin{equation}
H_{k\ell}
=
\sum_{a=1, a \neq k, a \neq \ell}^K
\left ( 
\sum_{b=1, b \neq a, b \neq k, b \neq \ell}^{a-1} \left ( \frac{1}{C_a} + \frac{1}{C_b} \right )
\right ).
\label{eq:Defn+H_kl}
\end{equation}

Collecting  terms  (\ref{eq:ZZ1})-(\ref{eq:ZZ5}), we conclude from (\ref{eq:ZZ0}) that
\begin{eqnarray}
& &
\bE{ \1{ \nu^\star_s = k} \1{s+ \frac{\sigma_s}{C_{k}} > 0} 
\1{\nu_t = \ell } \1{ t + \frac{\sigma_t}{C_{\ell}} > 0 } 
\frac{ \sigma_0 }{ C_{\nu_0}} }
\nonumber \\
&=&
\frac{ 1 }{K(K-1)} \cdot H^\star_{k\ell} \cdot 
\bP{ \nu^\star_s = k, s+ \frac{\sigma_s}{C_{k}} > 0, \nu_t = \ell , t + \frac{\sigma_t}{C_{\ell}} > 0 }
\cdot
\bE{ \sigma_0 }
\end{eqnarray}
with
\[
H^\star_{k\ell}
=
H_{k\ell}  + 4 \Gamma - 3 \left ( \frac{1}{C_k} + \frac{1}{C_\ell} \right ) .
\]

In Appendix D we show that
\begin{equation}
\bP{ \nu^\star_s = k, s+ \frac{\sigma_s}{C_{k}} > 0, \nu_t = \ell , t + \frac{\sigma_t}{C_{\ell}} > 0 }
= 
\frac{1}{K(K-1)} 
\bP{ s+ \frac{\sigma_s}{C_{k}} > 0 } \bP{t + \frac{\sigma_t}{C_{\ell}} > 0 }
\label{eq:ProbabilityZ}
\end{equation}
and the conclusion
\begin{eqnarray}
& &
\bE{ \1{ \nu^\star_s = k} \1{s+ \frac{\sigma_s}{C_{k}} > 0} 
\1{\nu_t = \ell } \1{ t + \frac{\sigma_t}{C_{\ell}} > 0 } 
\frac{ \sigma_0 }{ C_{\nu_0}} }
\nonumber \\
&=&
\frac{ H^\star_{k\ell} }{K^2(K-1)^2} 
\cdot 
\bP{ s+ \frac{\sigma_s}{C_{k}} > 0 } \bP{t + \frac{\sigma_t}{C_{\ell}} > 0 }
\cdot
\bE{ \sigma_0 }
\label{eq:ZZ6}
\end{eqnarray}
follows.
In Appendix E we also show that
\begin{equation}
H_{k\ell}
=
(K-3)
\left ( \Gamma -  \frac{1}{C_k} - \frac{1}{C_\ell} \right )
\label{eq:H_kell}
\end{equation}
so that
\begin{eqnarray}
H^\star_{k\ell}
&=&
H_{k\ell}  + 4 \Gamma - 3 \left ( \frac{1}{C_k} + \frac{1}{C_\ell} \right ) 
 \nonumber \\
 &=&
(K-3) \left ( \Gamma -  \frac{1}{C_k} - \frac{1}{C_\ell} \right )
+ 4 \Gamma - 3 \left ( \frac{1}{C_k} + \frac{1}{C_\ell} \right ) 
 \nonumber \\
 &=&
 (K+1) \Gamma - K \left ( \frac{1}{C_k} + \frac{1}{C_\ell} \right )
 = \Sigma_{k\ell}
\end{eqnarray}
with $\Sigma_{k\ell}$ given by (\ref{eq:Sigma_kl}).
Inserting this last expression into 
(\ref{eq:ZZ6}) yields the desired conclusion (\ref{eq:Case2B+2}).
\myendpf

\section*{Acknowledgements}
This research was partially carried out while the authors 
were in residence under the Saiotek Program on "Virtual Machines for the Traffic Analysis in High Capacity Networks" 
was partially supported by grant SA-2012/00331 (Department of Industry, 
Innovation, Trade and Tourism, Basque Government). 
The work of Ane Izagirre was also supported by the grants of the Ecole Doctorale EDSYS and INP.

\section*{References}

\bibliographystyle{plainnat}
\bibliography{bibli_PEVA16}


\section*{Appendix A: A proof of (\ref{eq:VAR})}
\label{App:A}

We are interested in assessing the range of values for
$\rm{Var}[X]$ under the constraint $\bE{X} = \frac{\Gamma}{K}$ for some $\Gamma > 0$.
This amounts to considering the expression
\[
\frac{1}{K} \sum_{k=1}^K \frac{1}{C^2_k} 
-
\left ( \frac{1}{K} \sum_{k=1}^K \frac{1}{C_k}  \right )^2,
\quad \myvec{C} = (C_1, \ldots , C_K) \in (0,\infty)^K
\]
under the constraint
\[
\sum_{k=1}^K \frac{1}{C_k} = \Gamma.
\]

Defining the mapping $g:  (0,\infty)^K \rightarrow \mathbb{R}_+ $ as
\[
g(\myvec{C}) \equiv \sum_{k=1}^K \frac{1}{C^2_k} ,
\quad \myvec{C} = (C_1, \ldots , C_K) \in (0,\infty)^K,
\]
we need only focus on studying the range of 
$ \left \{ g(\myvec{C}) : \  \myvec{C} \in \mathcal{C}(\Gamma) \right \} $
where the constraint set $\mathcal{C}(\Gamma)$ is given by
\[
\mathcal{C}(\Gamma)
= 
\left \{
\myvec{C} \in (0,\infty)^K:
\  \sum_{k=1}^K \frac{1}{C_k} = \Gamma
\right \} .
\]

This issue is more easily understood with the help of the change of variables
$T: (0,\infty)^K \rightarrow (0,\infty)^K$ given by
\[
T(\myvec{C} ) 
=
\left (
\frac{1}{C_1} , \ldots , \frac{1}{C_K} 
\right ),
\quad \myvec{C} = (C_1, \ldots , C_K) \in (0,\infty)^K.
\]
The transformation $T$ is a bijection from $(0,\infty)^K$ into itself  (with inverse $T^{-1} = T$).
Note that $T$ puts the set $\mathcal{C}(\Gamma)$ into one-to-one correspondence with the set $\mathcal{X}(\Gamma)$
given by
\[
\mathcal{X}(\Gamma)
= 
\left \{
\myvec{x} = (x_1, \ldots , x_K) \in (0,\infty)^K:
\  \sum_{k=1}^K x_k = \Gamma
\right \} .
\]
 
If we define the mapping $h:  \mathbb{R}_+^K \rightarrow \mathbb{R}_+ $ by
\[
h(\myvec{x}) \equiv  \sum_{k=1}^K x_k^2 ,
\quad \myvec{x} = (x_1, \ldots , x_K) \in \mathbb{R}_+^K ,
\]
then we obviously have
\begin{equation}
g(\myvec{C}) = h( T(\myvec{C})), 
\quad \myvec{C} \in (0,\infty)^K.
\label{eq:FromGtoH}
\end{equation}
Moreover it holds that
$
\left \{ g(\myvec{C}) : \  \myvec{C} \in \mathcal{C}(\Gamma) \right \} 
=
\left \{ h(\myvec{x}) : \  \myvec{x} \in \mathcal{X}(\Gamma) \right \}
$
since $T( \mathcal{C}(\Gamma) ) = \mathcal{X}(\Gamma)$.

With $\prec$ denoting majorization \citep[p. 7]{MarshallOlkin+Book},
whenever $\myvec{x}_1 \prec \myvec{x}_2$ in $\mathbb{R}_+^K$, we have
\begin{equation}
h(\myvec{x}_1) \leq  h( \myvec{x}_2 )
\label{eq:SchurConvexityInequality}
\end{equation}
by the Schur-convexity of the function $h: \mathbb{R}_+^K \rightarrow \mathbb{R}_+$
(inherited from the convexity of $t \rightarrow t^{2}$)
\citep[Prop. C.1, p. 64]{MarshallOlkin+Book}; see also \citep[p. 54]{MarshallOlkin+Book} for the definition of Schur-convexity. 
The most \lq\lq balanced" element of $\mathcal{X}(\Gamma)$ is the vector 
$\myvec{x}_\star$ given by 
\[
\myvec{x}_\star = \frac{\Gamma}{K} \left ( 1, \ldots , 1 \right ).
\]
It represents the \lq\lq smallest" element in the constraint set
in the sense of majorization \citep[p. 7]{MarshallOlkin+Book} -- We have
$ \myvec{x}_\star \prec \myvec{x}$  for any $\myvec{x}$ in $\mathcal{X}(\Gamma)$, whence
$h(\myvec{x}_\star) \leq  h( \myvec{x} )$ by (\ref{eq:SchurConvexityInequality}).
With $\myvec{C}_\star$ given by (\ref{eq:BalancedCapacities}) we see from (\ref{eq:FromGtoH}) that
\[
g( \myvec{C}_\star) \leq g( \myvec{C}),
\quad \myvec{C} \in \mathcal{C}(\Gamma) 
\]
since $\myvec{C}_\star = T( \myvec{x}_\star)$.
This establishes the lower bound  in (\ref{eq:VAR}) in agreement with the earlier discussion
concerning the zero variance when all the capacities are identical.

We now turn to the upper bound: For each $k=1, \ldots , K$, introduce the vector $\myvec{x}^\star_k$ in $\mathbb{R}_+^K$ given by
\[
\myvec{x}^\star_k = \Gamma \myvec{e}_k
\]
with $\myvec{e}_1, \ldots , \myvec{e}_K$ as defined in Section \ref{sec:Discussion}.
It is well known \citep[p. 7]{MarshallOlkin+Book} that
$\myvec{x} \prec \myvec{x}^\star_k$ for every 
$\myvec{x}$ in $ \mathcal{X}(\Gamma) $, so that
$h(\myvec{x}) \leq h(\myvec{x}^\star_k)$ for every 
$\myvec{x}$ in $ \mathcal{X}(\Gamma) $.
Although $\myvec{x}^\star_k$ is {\em not} an element of $\mathcal{X}(\Gamma) $, we nevertheless have 
\[
\sup \left \{ h(\myvec{x}) : \  \myvec{x} \in \mathcal{X}(\Gamma) \right \}
=
\sup \left \{ h(\myvec{x}) : \  \myvec{x} \in \overline{ \mathcal{X}(\Gamma) } \right \}
= h(\myvec{x}^\star _k) 
\]
by the continuity of $h$;  the closure $\overline{ \mathcal{X}(\Gamma) } $  of $\mathcal{X}(\Gamma)$
is given by
$ \overline{ \mathcal{X}(\Gamma) } = \left \{ \myvec{x} \in \mathbb{R}_+^K : \ \sum_{k=1}^K x_k = \Gamma \right \}$.
In particular, we have
\begin{eqnarray}
\sup \left \{ g(\myvec{C}) : \  \myvec{C} \in \mathcal{C}(\Gamma) \right \}
=
h(\myvec{x}^\star _k) 
=
\Gamma^2.
\end{eqnarray}

Note that for each $k=1, \ldots , K$ there is no vector $\myvec{C}^\star_k$ in $\mathcal{C}(\Gamma) $ such that
$\myvec{x}^\star_k = T( \myvec{C}^\star_k )$.
However, there are vectors in $\mathcal{C}(\Gamma) $ whose value under $g$ will come arbitrarily close to $\Gamma^2$.
For instance, consider the vectors
\[
\myvec{x}_{k,a} 
= 
a\Gamma \myvec{e}_k 
+  \frac{(1-a)\Gamma}{K-1} \sum_{\ell=1, \ell \neq k }^K \myvec{e}_\ell ,
\quad 
\begin{array}{c}
0 < a < 1 \\
k=1, \ldots , K. \\
\end{array}
\]
These vectors are elements of $\mathcal{X}(\Gamma) $ with
$\lim_{a \uparrow 1 } \myvec{x}_{k,a} = \myvec{x}^\star_k$, hence
$\lim_{a \uparrow 1 } h( \myvec{x}_{k,a} ) = h( \myvec{x}^\star_k)$ by continuity.
As we recall the definition (\ref{eq:C_ak}) we check that
\[
\myvec{C}_{k,a} 
=
\frac{1}{a\Gamma} \myvec{e}_k 
+  \frac{K-1}{(1-a)\Gamma} \sum_{\ell=1, \ell \neq k }^K \myvec{e}_\ell 
=
T( \myvec{x}_{k,a}  ),
\quad
\begin{array}{c}
0 < a < 1 \\
k=1, \ldots , K \\
\end{array}
\]
so that $h( \myvec{x}_{k,a}  ) = g( \myvec{C}_{k,a} )$ for all $0 < a < 1$. It follows that
\[
\lim_{a \uparrow 1 }  g( \myvec{C}_{k,a} ) 
=
\lim_{a \uparrow 1 }  h( \myvec{x}_{k,a}  ) 
=
h( \myvec{x}^\star_k)
= \Gamma^2
\]
and this completes the discussion of the upper bound at (\ref{eq:VAR}).
\myendpf

\section*{Appendix B: A proof of (\ref{eq:Sigma_k})}
\label{App:B}

Fix $k=1,2, \ldots , K$.
Elementary calculations give
\begin{eqnarray}
& & 
\sum_{ a=1 , a \neq k }^K
\left (\sum_{b=1, b \neq k}^{a-1}
\left ( \frac{1}{C_a} + \frac{1}{C_b}  \right )
\right )
\nonumber \\
&=&
\sum_{a=1}^{k-1}
\left (\sum_{b=1, b \neq k}^{a-1}
\left ( \frac{1}{C_a} + \frac{1}{C_b}  \right )
\right )
+
\sum_{a=k+1}^K
\left (\sum_{b=1, b \neq k}^{a-1}
\left ( \frac{1}{C_a} + \frac{1}{C_b}  \right )
\right )
\nonumber \\
&=&
\sum_{a=1}^{k-1}
\left (\sum_{b=1}^{a-1}
\left ( \frac{1}{C_a} + \frac{1}{C_b}  \right )
\right )
+
\sum_{a=k+1}^K
\left ( - \frac{1}{C_a} - \frac{1}{C_k}
+ \sum_{b=1}^{a-1}
\left ( \frac{1}{C_a} + \frac{1}{C_b}  \right )
\right )
\nonumber \\
&=&
\sum_{a=1}^{k-1} \frac{a-1}{C_a}
+ \sum_{a=k+1}^K \frac{a-1}{C_a}
+ \sum_{a=1}^{k-1} \left ( \sum_{b=1}^{a-1} \frac{1}{C_b} \right )
+ \sum_{a=k+1}^K \left ( 
- \frac{1}{C_a} - \frac{1}{C_k}
+ \sum_{b=1}^{a-1} \frac{1}{C_b} \right )
\nonumber \\
&=& 
\sum_{a=1}^{K} \frac{a-1}{C_a}
- \frac{k-1}{C_k}
+ \sum_{a=1}^{K} \left ( \sum_{b=1}^{a-1} \frac{1}{C_b} \right )
- \sum_{b=1}^{k-1} \frac{1}{C_b}
- \sum_{a=k+1}^K \left (\frac{1}{C_a} + \frac{1}{C_k} \right )
\nonumber \\
&=&
\sum_{a=1}^{K} \frac{a-1}{C_a}
- \frac{k-1}{C_k}
+ \sum_{a=1}^{K} \left ( \sum_{b=1}^{a-1} \frac{1}{C_b} \right )
- \sum_{a=1}^{K} \frac{1}{C_a}
- \frac{K-k-1}{C_k}
\nonumber \\
&=&
\sum_{a=1}^{K} \frac{a-1}{C_a}
- \frac{K-2}{C_k}
+ \sum_{a=1}^{K} \left ( \sum_{b=1}^{a-1} \frac{1}{C_b} \right )
- \sum_{a=1}^{K} \frac{1}{C_a}
\nonumber \\
&=&
\sum_{a=1}^{K} \frac{a-1}{C_a}
- \frac{K-2}{C_k}
+ \sum_{b=1}^{K-1} \left ( \sum_{a=b+1}^{K} \frac{1}{C_b} \right )
- \sum_{a=1}^{K} \frac{1}{C_a}
\nonumber \\
&=&
\sum_{a=1}^{K} \frac{a-1}{C_a}
- \frac{K-2}{C_k}
+ \sum_{b=1}^{K-1} \frac{K-b}{C_b}
- \sum_{a=1}^{K} \frac{1}{C_a}
\nonumber \\
&=&
(K-1) \Gamma - \frac{K-2}{C_k} - \Gamma 
\end{eqnarray}
and the proof of (\ref{eq:Sigma_k}) is complete.
\myendpf

\section*{Appendix C: A proof of (\ref{eq:AppB+1})}
\label{App:C}

We are in the situation $s < t < 0$. Fix $k=1, \ldots , K$.
Our point of departure is the obvious decomposition
\begin{equation}
\bP{ t < s + \frac{\sigma_s}{C_{\nu^\star_s}} \leq 0, \nu_t  = k }
=
\sum_{\ell=1, \ell \neq k}^K
\bP{ \nu^\star_s = \ell, t < s + \frac{\sigma_s}{C_\ell} \leq 0, \nu_t  = k} .
\label{eq:PieceZ1}
\end{equation} 
Pick $\ell =1, \ldots, K$ distinct from $k$, and note that
\begin{eqnarray}
\bP{ \nu^\star_s = \ell, t < s + \frac{\sigma_s}{C_\ell} \leq 0, \nu_t  = k}
&=&
\bP{ \nu^\star_s = \ell, t < s + \frac{\sigma_s}{C_\ell} \leq 0, 
\ell \in \Sigma_t, \nu_t  = k}
\nonumber \\
& & ~+
\bP{ \nu^\star_s = \ell, t < s + \frac{\sigma_s}{C_\ell} \leq 0, 
\ell \notin \Sigma_t, \nu_t  = k}.
\label{eq:PieceZ2}
\end{eqnarray}

We examine each term in turn: 
First, when $\ell$ belongs to $\Sigma_t$ with $\nu^\star_s = \ell$, then $\nu_t  = k$
happens only if $\nu^\star_s = \ell $ and $\Sigma_t = \{k,\ell \}$,
whence
\begin{eqnarray}
\bP{ \nu^\star_s = \ell, t < s + \frac{\sigma_s}{C_\ell} \leq 0, 
\ell \in \Sigma_t, \nu_t  = k}
&=&
\bP{ \nu^\star_s = \ell, t < s + \frac{\sigma_s}{C_\ell} \leq 0, 
\Sigma_t = \{ k,\ell \}, \nu_t  = k}
\nonumber \\
&=&
\bP{ \nu^\star_s = \ell, t < s + \frac{\sigma_s}{C_\ell} \leq 0, 
\Sigma_t = \{ k,\ell \}}
\nonumber \\
&=&
\frac{2}{K^2(K-1)} \bP{ t < s + \frac{\sigma_s}{C_\ell} \leq 0}.
\label{eq:PieceX}
\end{eqnarray}
Next, we have $\nu_t = \nu^\star_t$ when $\nu^\star_s$ is not in $\Sigma_t$, 
so that
\begin{eqnarray}
\lefteqn{
\bP{ \nu^\star_s = \ell, t < s + \frac{\sigma_s}{C_\ell} \leq 0,
\ell \notin \Sigma_t, \nu_t  = k}
} & &
\nonumber \\
&=&
\sum_{a=1, a\neq k, a \neq \ell}^{K}
\bP{ \nu^\star_s = \ell, t < s + \frac{\sigma_s}{C_\ell} \leq 0,
\ell \notin \Sigma_t, \Sigma_t = \{k,a\}, \nu^\star_t  = k}
\nonumber \\
&=&
\frac{1}{2} \sum_{a=1, a\neq k, a \neq \ell}^{K}
\bP{ \nu^\star_s = \ell, t < s + \frac{\sigma_s}{C_\ell} \leq 0,
\Sigma_t = \{k,a\} }
\nonumber \\
&=&
\frac{1}{2} \sum_{a=1, a\neq k, a \neq \ell}^{K}
\frac{1}{K} \frac{2}{K(K-1)} \bP{ t < s + \frac{\sigma_s}{C_\ell} \leq 0 }
\nonumber \\
&=&
\frac{1}{K^2(K-1)} 
\sum_{a=1, a\neq k, a \neq \ell}^{K}
\bP{ t < s + \frac{\sigma_s}{C_\ell} \leq 0 }
\nonumber \\
&=&
\frac{K-2}{K^2(K-1)} 
\bP{ t < s + \frac{\sigma_s}{C_\ell} \leq 0 } .
\label{eq:PieceY}
\end{eqnarray}

Inserting (\ref{eq:PieceX}) and (\ref{eq:PieceY})
back into (\ref{eq:PieceZ2}) we get
\begin{equation}
\bP{ \nu^\star_s = \ell, t < s + \frac{\sigma_s}{C_\ell} \leq 0, \nu_t  = k}
=
\frac{1}{K(K-1)} \bP{ t < s + \frac{\sigma_s}{C_\ell} \leq 0 }
\label{eq:PieceZ2withX+Y}
\end{equation}
and the desired conclusion (\ref{eq:AppB+1})
follows with the help of (\ref{eq:PieceZ1}).
\myendpf



\section*{Appendix D: A proof of (\ref{eq:ProbabilityZ})}
\label{App:D}

Recall that we are in the situation $s < t < 0$.  
Fix distinct $k,\ell =1, \ldots , K$. We need to show that
\begin{equation}
\bP{ \nu^\star_s = k, s+ \frac{\sigma_s}{C_{k}} > 0, \nu_t = \ell , t + \frac{\sigma_t}{C_{\ell}} > 0 }
= 
\frac{1}{K(K-1)} 
\bP{ s+ \frac{\sigma_s}{C_{k}} > 0 } \bP{t + \frac{\sigma_t}{C_{\ell}} > 0 }
\label{eq:ProbabilityZAgain}
\end{equation}

By arguments used earlier we get
\begin{eqnarray}
& & 
\bP{ \nu^\star_s = k, s+ \frac{\sigma_s}{C_{k}} > 0, k \in \Sigma_t , \nu_t = \ell , t + \frac{\sigma_t}{C_{\ell}} > 0 }
\nonumber \\
&=&
\bP{ \nu^\star_s = k, s+ \frac{\sigma_s}{C_{k}} > 0,  \Sigma_t = \{ k, \ell \} , \nu_t = \ell , t + \frac{\sigma_t}{C_{\ell}} > 0 }
\nonumber \\
&=&
\bP{ \nu^\star_s = k, s+ \frac{\sigma_s}{C_{k}} > 0,  \Sigma_t = \{ k, \ell \} , t + \frac{\sigma_t}{C_{\ell}} > 0 }
\nonumber \\
&=&
\frac{2}{K^2(K-1)} 
\bP{ s+ \frac{\sigma_s}{C_{k}} > 0 } \bP{t + \frac{\sigma_t}{C_{\ell}} > 0 }
\label{eq:ProbabilityZZZZ1}
\end{eqnarray}
under the enforced independence assumptions.

In a similar way, we find
\begin{eqnarray}
& &
\bP{ \nu^\star_s = k, s+ \frac{\sigma_s}{C_{k}} > 0, k \notin \Sigma_t ,  \nu_t = \ell , t + \frac{\sigma_t}{C_{\ell}} > 0 }
\nonumber \\
&=&
\sum_{ a=1, a \neq k, a \neq \ell}^K
\bP{ \nu^\star_s = k, s+ \frac{\sigma_s}{C_{k}} > 0,  \Sigma_t = \{ a, \ell \} , \nu_t = \ell , t + \frac{\sigma_t}{C_{\ell}} > 0 }
\nonumber \\
&=&
\frac{1}{2}
\sum_{ a=1, a \neq k, a \neq \ell}^K
\bP{ \nu^\star_s = k, s+ \frac{\sigma_s}{C_{k}} > 0,  \Sigma_t = \{ a, \ell \} , t + \frac{\sigma_t}{C_{\ell}} > 0 }
\nonumber \\
&=&
\frac{1}{K^2(K-1)} 
\sum_{ a=1, a \neq k, a \neq \ell}^K
\bP{ s+ \frac{\sigma_s}{C_{k}} > 0 } \bP{t + \frac{\sigma_t}{C_{\ell}} > 0 }
\nonumber \\
&=&
\frac{K-2}{K^2(K-1)} 
\bP{ s+ \frac{\sigma_s}{C_{k}} > 0 } \bP{t + \frac{\sigma_t}{C_{\ell}} > 0 }
\label{eq:ProbabilityZZZZ2}
\end{eqnarray}
under the enforced independence assumptions.
Collecting (\ref{eq:ProbabilityZZZZ1}) and (\ref{eq:ProbabilityZZZZ2}) we conclude to the validity of (\ref{eq:ProbabilityZAgain}).
\myendpf

\section*{Appendix E: A proof of (\ref{eq:H_kell})}
\label{App:E}

To show (\ref{eq:H_kell})
it suffices to establish this fact for $k=1$ and $\ell=2$ -- This follows from the fact that the labeling of the servers
is arbitrary.
Thus, form the definition (\ref{eq:Defn+H_kl}) we get
\begin{eqnarray}
H_{12}
&=&
\sum_{a=3}^K
\sum_{b=3}^{a-1} \left ( \frac{1}{C_a} + \frac{1}{C_b} \right )
\nonumber \\
&=&
\sum_{a=3}^K
\left (
\sum_{b=1}^{a-1} \left ( \frac{1}{C_a} + \frac{1}{C_b} \right )
- \left (  \frac{1}{C_a} + \frac{1}{C_1} + \frac{1}{C_a} + \frac{1}{C_2} \right )
\right )
\nonumber \\
&=&
\sum_{a=3}^K
\left (
\frac{a-3}{C_a} + \sum_{b=1}^{a-1} \frac{1}{C_b}  - \frac{1}{C_1} - \frac{1}{C_2} 
\right )
\nonumber \\
&=&
\sum_{a=3}^K \frac{a-3}{C_a} 
+ \sum_{a=3}^K \left ( \sum_{b=1}^{a-1} \frac{1}{C_b}  \right )
- (K-2) \left ( \frac{1}{C_1} + \frac{1}{C_2} \right )
\nonumber \\
&=&
\sum_{a=3}^K \frac{a-3}{C_a} 
+ \sum_{a=1}^K \left ( \sum_{b=1}^{a-1} \frac{1}{C_b}  \right )
- (K-2) \left ( \frac{1}{C_1} + \frac{1}{C_2} \right )
- \frac{1}{C_1}
\nonumber \\
&=&
\sum_{a=1}^K \frac{a-3}{C_a} 
+ \sum_{a=1}^K \left ( \sum_{b=1}^{a-1} \frac{1}{C_b}  \right )
- (K-2) \left ( \frac{1}{C_1} + \frac{1}{C_2} \right )
- \frac{1}{C_1}
- \left ( - \frac{2}{C_1} - \frac{1}{C_2}
\right )
\nonumber \\
&=&
\sum_{a=1}^K \frac{a-3}{C_a} 
+ \sum_{a=1}^K \left ( \sum_{b=1}^{a-1} \frac{1}{C_b}  \right )
- \frac{K-3}{C_1} - \frac{K-3}{C_2}
\nonumber \\
&=&
\sum_{a=1}^K \frac{a-3}{C_a} 
+ \sum_{b=1}^{K-1} \left ( \sum_{a=b+1}^{K} \frac{1}{C_b}  \right )
- \frac{K-3}{C_1} - \frac{K-3}{C_2}
\nonumber \\
&=&
\sum_{a=1}^K \frac{a-3}{C_a}  + \sum_{b=1}^{K-1}  \frac{K-b}{C_b} 
- \frac{K-3}{C_1} - \frac{K-3}{C_2}
\nonumber \\
&=&
\sum_{a=1}^{K-1} \frac{K-3}{C_a}  + \frac{K-3}{C_K} 
- \frac{K-3}{C_1} - \frac{K-3}{C_2},
\nonumber
\end{eqnarray}
whence (\ref{eq:H_kell}) follows as announced.
\myendpf
\end{document}